\documentclass{iopart}

\usepackage{graphicx}
\usepackage{xcolor}
\usepackage{hyperref}

\begin{document}


\title{Sliding Dynamics of Skyrmion Molecular Crystals}

\author{J. C. Bellizotti Souza$^{1, *}$,
  C. J. O. Reichhardt$^2$,
  C. Reichhardt$^2$,
  N. P. Vizarim$^3$ and
  P. A. Venegas$^4$}

\ead{jc.souza@unesp.br}
\address{$^*$ Corresponding author}
\address{$^1$ POSMAT - Programa de P\'os-Gradua\c{c}\~ao em Ci\^encia
  e Tecnologia de Materiais, S\~ao Paulo State University (UNESP),
  School of Sciences, Bauru 17033-360, SP, Brazil}

\address{$^2$ Theoretical Division and Center for Nonlinear Studies,
  Los Alamos National Laboratory, Los Alamos, New Mexico 87545, USA}

\address{$^3$ “Gleb Wataghin” Institute of Physics, University of
  Campinas, 13083-859 Campinas, S\~ao Paulo, Brazil}

\address{$^4$ Department of Physics, S\~ao Paulo State University
  (UNESP), School of Sciences, Bauru 17033-360, SP, Brazil}

\date{\today}

\begin{abstract}
Using both atomistic and particle-based simulations, we investigate the current-driven dynamics of skyrmions on two-dimensional periodic substrates when there are multiple skyrmions per substrate minimum. At zero drive, the system forms pinned skyrmion molecular crystal states consisting of dimers, trimers, or dimer-trimer mixtures that have both positional and orientational order. On a square substrate lattice, the motion above depinning occurs via a running soliton that travels completely transverse to the applied current. This motion is generated by a torque from the Magnus force, which rotates the $n$-mer states perpendicular to the applied current. At higher drives, the flow becomes disordered while the Hall angle diminishes and gradually approaches the intrinsic value. In some cases, we also find directional locking where the Hall angle becomes locked to certain symmetry directions of the substrate over a range of currents. The transitions into and out of directionally locked states are accompanied by negative differential mobility in which the net velocity decreases as the drive increases. On a triangular substrate, we find no transverse mobility effects, but still observe directionally locked motion.
\end{abstract}

\maketitle

\section{Introduction}

There are a variety of systems that can be effectively described as
interacting particles coupled to a two-dimensional periodic
substrate, and in these systems, the interparticle interactions
favor a certain spacing that may differ from the energy-minimizing spacing
imposed by the substrate.
Under these conditions,
commensurate-incommensurate transitions can occur as the ratio
of the number of particles to the number of substrate minima is varied
\cite{Bak82,Coppersmith82,Harada96,Bohlein12,Reichhardt17}.
For commensurate conditions, the number of particles
is an integer multiple of the number of substrate
minima, such as a one-to-one matching.
At incommensurate fillings, the system still
forms an ordered structure but
contains interstitials or
vacancies that act as kinks or antikinks \cite{Bak82,Bohlein12}.
In
some cases, the density of incommensurations is high enough to cause the
system to disorder.
Ordered commensurate states can also appear for
fractional fillings such as $1/2$ or $1/3$
\cite{Hallen93,Reichhardt01,Field02,Grigorenko03}. Under an
applied drive, the threshold force needed to depin the particles is
maximized at the commensurate conditions, and a single
depinning transition occurs. At incommensurate fillings, the depinning
threshold can be dramatically reduced and can also
occur via a multistep process in which
the kinks or antikinks begin to slide first, followed by the sliding
of the remaining particles at higher drives \cite{Reichhardt17}.

Commensurate-incommensurate transitions have been studied
for the ordering of
atoms and molecules on periodic surfaces \cite{Bak82,Coppersmith82},
vortices in type-II superconductors with artificial periodic pinning
arrays \cite{Baert95,Harada96,Reichhardt01,Field02,Grigorenko03},
vortices in Bose-Einstein condensates on optical trap arrays
\cite{tung_observation_2006}, charged colloids on optical traps or
patterned substrates
\cite{brunner_phase_2002,Tierno07,Mikhael08,Bohlein12}, dusty plasmas
\cite{Zhu22}, liquid crystal skyrmions on ordered substrates
\cite{duzgun_commensurate_2020} and generalized Wigner crystals for
charge ordering on periodic substrates in
moir{\' e} systems \cite{Xu20,Wigner25}.
Most studies of commensurate-incommensurate systems have been performed
for one or slightly more than one particle per substrate minimum.
In superconducting vortex
systems, at commensurate fillings for which there are two or more vortices
per trap, the individual vortices can merge to form multiquantum vortices
\cite{Baert95}. There can also be situations
for which the trapped particles are unable to merge,
and the system contains $n$ separate particles per trap. In this case,
each substrate minimum contains an $n$-mer state that has positional
ordering produced by the substrate periodicity.
The $n$-mers may also exhibit
an additional orientational ordering with the other $n$-mers
\cite{reichhardt_novel_2002,brunner_phase_2002,agra_theory_2004,frey_melting_2005,reichhardt_ordering_2005,thomas_spin_2007,reichhardt_vortex_2007,neal_competing_2007}.

When multiple colloidal particles are trapped by each substrate potential
minimum in an optical trap array,
the ordered states are called colloidal molecular crystals
\cite{reichhardt_novel_2002,brunner_phase_2002,agra_theory_2004,frey_melting_2005}.
For filings $N = 2.0$, the colloids form dimers that can arrange
into herringbone lattices where the tilt angle of the dimers alternates
from one row to the next,
antiferromagnetic lattices in which adjacent dimers are
perpendicular to each other, or ferromagnetic states in which all of the
dimers have the same alignment. Trimers can also exhibit ferromagnetic and
columnar ordering depending on the
substrate symmetry and strength
\cite{reichhardt_novel_2002,brunner_phase_2002,agra_theory_2004,frey_melting_2005}.
As a function of increasing temperature or changing colloidal
interaction strength, the molecular
crystals can undergo a transition from an orientationally ordered
to an orientationally disordered state in which each substrate minimum
still contains $n$ particles that form an $n$-mer, but the
global orientational ordering of the
$n$-mers is lost
\cite{reichhardt_novel_2002,brunner_phase_2002,agra_theory_2004,frey_melting_2005}.
At higher temperatures, the system forms a modulated liquid  in which
colloidal particles can hop among the substrate minima
\cite{reichhardt_novel_2002,brunner_phase_2002,mikulis_re-entrant_2004,reichhardt_ordering_2005}.
More recently, Wigner crystal molecule crystals were
observed for charge ordering of two
or more electrons per trap on ordered substrates \cite{Li24}.

There have also been studies of colloidal molecular crystal mixtures
at fillings of $3/2$ and $5/2$ in which monomer-dimer or
dimer-trimer mixtures are present that
create larger scale superlattice
orderings \cite{reichhardt_ordering_2005}.
Recently, magnetic
skyrmion molecular crystals were proposed for skyrmions interacting
with periodic two-dimensional substrates when two or more
skyrmions per substrate minimum are present
\cite{Souza25}. Such states could
arise for skyrmions in moir\'{e} materials
\cite{hejazi_heterobilayer_2021} or skyrmions in systems with ordered
nanostructures, such as from patterned irradiation or surface
modulation \cite{juge_helium_2021}. Skyrmion molecular crystals are
distinct from particle-based molecular crystals in that
the skyrmions can
distort or annihilate,
making it possible for ordered lattices
to occur in which the skyrmions have
different sizes in the same trap, are stretched into meron pairs,
or
partially annihilate to make the state more
ordered \cite{Souza25}.

In particle-based systems, driven colloidal molecular crystals
have been shown to exhibit a variety of moving phases, and transitions
between
the different flow phases
can be correlated with features in the velocity-force curves
\cite{Reichhardt09,Reichhardt12}. The dynamic states
include plastic flow phases, where some particles move while others
remain pinned, fluid-like flows, and moving crystal phases.
An open question is whether the dynamics of the deformable
skyrmion molecular
crystal states are similar or different
from the dynamics observed for rigid colloidal molecular crystal
states.

Skyrmions are particle-like magnetic textures that
can form a triangular lattice
\cite{muhlbauer_skyrmion_2009,yu_real-space_2010,nagaosa_topological_2013},
and be set into motion by an
applied spin current \cite{jonietz_spin_2010,iwasaki_universal_2013}.
When they interact with defects or interfaces, pinning effects
can arise and there can be a finite
threshold current for motion \cite{reichhardt_statics_2022}. A key feature
that distinguishes skyrmion dynamics from the dynamics of
most other systems is that skyrmions have a strong non-dissipative Magnus force
component to their motion
\cite{nagaosa_topological_2013,everschor-sitte_perspective_2018,reichhardt_statics_2022}.
The Magnus force causes skyrmions to have a finite Hall angle with respect
to an external drive and also affects how the skyrmions interact
with interfaces, pinning, and other skyrmions
\cite{jiang_direct_2016,litzius_skyrmion_2016,brearton_deriving_2021,reichhardt_statics_2022}.
For skyrmions interacting with pinning sites, the Magnus force generates
a spiraling motion that can reduce the pinning effectiveness or
modify the Hall angle
\cite{reichhardt_thermal_2018,reichhardt_statics_2022}.
The skyrmion
Hall angle becomes strongly drive dependent in the presence of pinning,
starting from a value of nearly zero at
low drives just above depinning and then increasing
as the drive increases
\cite{reichhardt_nonequilibrium_2018,jiang_direct_2016,litzius_skyrmion_2016}.
The pinning reduces the Hall angle by inducing
a side jump of the skyrmions as
they move over the pinning sites, and the size of this side jump
diminishes with increasing drive
so that at high drives, the skyrmion travels along a direction
close to the intrinsic skyrmion Hall angle
\cite{reichhardt_statics_2022}.
In a particle-based model,
the motion of a single skyrmion driven with an increasing current
over a periodic substrate shows a directional locking effect in which
the skyrmion Hall angle increases in a series
of steps as the skyrmion motion becomes locked to specific symmetry
directions of the substrate over an interval of drives.
For a square array, the motion is most strongly locked along
$45^\circ$, but weaker locking occurs
for other angles $ \theta = \arctan(n/m)$, where
$n$ and $m$ are integers.
For a triangular substrate, the strongest
directional locking occurs along
$60^\circ$
\cite{Reichhardt15a,vizarim_skyrmion_2020}. This directional locking
effect also occurs for overdamped particles such as vortices and colloids
moving over a periodic substrate; however, in those systems the direction,
and not merely the magnitude,
of the drive must be varied
\cite{Reichhardt99,Korda02,MacDonald03,cao_orientational_2019}.

In studies of skyrmions driven over a
quasi-one-dimensional substrate when the number of skyrmions is just
above or below the first matching filling \cite{vizarim_shapiro_2020},
the depinning occurs through the motion of soliton kink at a
lower drive. The other skyrmions depin at a higher drive,
leading to a two-step depinning process. For skyrmions on
two-dimensional
substrates just above or below the first matching
filling, depinning also
occurs by kink or anti-kink flow in which directional locking
of the kinks is possible
\cite{Souza24}. Both
atomistic and particle-based models
of this process produce the same dynamic phases, but give
quantitative differences due to the ability of the skyrmions to distort
only in the atomistic model.

Here, we use atomistic and particle-based simulations to
examine the dynamics of skyrmion molecular crystals
for fillings of $N_\mathrm{sk}/N_m = 3/2$, 2.0, $5/2$, and
$3.0$ on square and triangular substrates, where
$N_\mathrm{sk}$ is the number of skyrmions and $N_m$ is the number of
substrate minima.
Previous work explored only the static ordering of
skyrmions on triangular substrates \cite{Souza25}.
For square
substrates at a filling of $N_\mathrm{sk}/N_m = 2.0$,
the system forms dimers with
antiferromagnetic ordering, similar to the ordering found in colloidal
molecular crystals on square lattices
\cite{reichhardt_novel_2002,agra_theory_2004}.
For $N_\mathrm{sk}/N_m = 3.0$, trimers appear that
form a columnar ordered phase,
while the $N_\mathrm{sk}/N_m = 1.5$ and $N_\mathrm{sk}/N_m = 2.5$ systems form
monomer-dimer and dimer-trimer mixtures, respectively.
We consider a
system with an intrinsic Hall angle of $\theta_\mathrm{int} = 26^\circ$.
For the
square substrates, depinning of the dimers and trimers occurs into a
state where the
flow is perpendicular to the current with a Hall angle
of $\theta=90^\circ$. At
higher drives, the Hall angle gradually decreases toward
$\theta_\mathrm{int}$.
We also find directional locking to $45^\circ$,
one of the symmetry directions of the
square substrate.
The transitions into and out of the directionally
locked states are associated with
negative differential mobility
in which the net velocity decreases with
increasing drive \cite{reichhardt_statics_2022}.
For a filling of $N_\mathrm{sk}/N_m = 3/2$,
where a mixture of dimers and monomers is present,
the absolute transverse mobility is lost, but there is strong
directional locking along $45^\circ$ in which solitons flow
along rows with dimers while monomer rows remain pinned.
At $N_\mathrm{sk}/N_m = 5/2$ where there is a dimer-trimer mixture,
we find a region of absolute transverse mobility accompanied at each end
by negative differential mobility.
For triangular substrates,
absolute transverse motion does not occur,
but motion above the initial depinning
transition runs close to $60^\circ$
along a substrate symmetry direction.
Generally, the depinning threshold is much higher for triangular substrates
than for square substrates.
We observe the same phases
with a particle-based model but find some
quantitative differences between the two models.

\section{Methods}

\subsection{Atomistic simulations}

Atomistic simulations capture the dynamics of
individual atomic magnetic moments \cite{evans_atomistic_2018}. We
model a ferromagnetic ultrathin film capable of holding
N{\'e}el skyrmions. Our sample has dimensions of 84 nm $\times$ 84 nm with
periodic boundary conditions along the $x$ and $y$ directions. We apply a
magnetic field perpendicular to the sample along the $-z$
direction and work at zero
temperature, $T=0$ K.

The Hamiltonian governing the atomistic dynamics is given by
\cite{iwasaki_universal_2013,evans_atomistic_2018,iwasaki_current-induced_2013}:

\begin{eqnarray}\label{eq:1}
  \mathcal{H}=&-\sum_{\langle i,
    j\rangle}J_{ij}\mathbf{m}_i\cdot\mathbf{m}_j -\sum_{\langle i,
    j\rangle}\mathbf{D}_{ij}\cdot\left(\mathbf{m}_i\times\mathbf{m}_j\right)\\\nonumber
  &-\sum_i\mu\mathbf{H}\cdot\mathbf{m}_i -\sum_{i} K(x_i,
  y_i)\left(\mathbf{m}_i\cdot\hat{\mathbf{z}}\right)^2 \ . \\\nonumber
\end{eqnarray}

The ultrathin film is modeled as a square arrangement of atoms with a
lattice constant $a=0.5$ nm. The first term on the right hand side is
the exchange interaction with an exchange constant of $J_{ij}=J$
between magnetic moments $i$ and $j$. The second term is the
interfacial Dzyaloshinskii–Moriya interaction, where
$\mathbf{D}_{ij}=D\mathbf{\hat{z}}\times\mathbf{\hat{r}}_{ij}$ is the
Dzyaloshinskii–Moriya vector between magnetic moments $i$ and $j$ and
$\mathbf{\hat{r}}_{ij}$ is the unit distance vector between sites $i$
and $j$. Here, $\langle i, j\rangle$ indicates that the sum is over
only the first neighbors of the $i$th magnetic moment. The third term
is the Zeeman interaction with an applied external magnetic field
$\mathbf{H}$. Here $\mu=g\mu_B$ is the magnitude of the magnetic
moment, $g=|g_e|=2.002$ is the electron $g$-factor, and
$\mu_B=9.27\times10^{-24}$~J~T$^{-1}$ is the Bohr magneton. The last
term represents the sample perpendicular magnetic anisotropy (PMA),
where $x_i$ and $y_i$ are the spatial coordinates of the $i$th
magnetic moment. In ultrathin films, long-range dipolar interactions
act as a PMA (see Supplemental of Wang {\it et al.} \cite{wang_theory_2018}),
and therefore merely effectively shift the PMA values.

The time evolution of atomic magnetic moments is obtained using the
Landau-Lifshitz-Gilbert (LLG)
equation \cite{seki_skyrmions_2016,gilbert_phenomenological_2004}:

\begin{equation}\label{eq:2}
  \frac{\partial\mathbf{m}_i}{\partial
    t}=-\gamma\mathbf{m}_i\times\mathbf{H}^\mathrm{eff}_i
  +\alpha\mathbf{m}_i\times\frac{\partial\mathbf{m}_i}{\partial t}
  +\frac{pa^3}{2e}\left(\mathbf{j}\cdot\nabla\right)\mathbf{m}_i \ .
\end{equation}
Here $\gamma=1.76\times10^{11}~$T$^{-1}$~s$^{-1}$ is the electron
gyromagnetic ratio,
$\mathbf{H}^\mathrm{eff}_i=-\frac{1}{\mu}\frac{\partial \mathcal{H}}{\partial \mathbf{m}_i}$
is the effective magnetic field including all interactions from the Hamiltonian, $\alpha$ is the
phenomenological damping introduced by Gilbert, and the last term is
the adiabatic spin-transfer-torque (STT) caused by application of an in
plane spin polarized current, where $p$ is the spin polarization, $e$
the electron charge, and $\mathbf{j}=j\hat{\mathbf{x}}$ the applied
current density. Use of this STT expression implies that the
conduction electron spins are always parallel to the magnetic moments
$\mathbf{m}$\cite{iwasaki_universal_2013,zang_dynamics_2011}.

\begin{figure}
  \centering
  \includegraphics[width=\columnwidth]{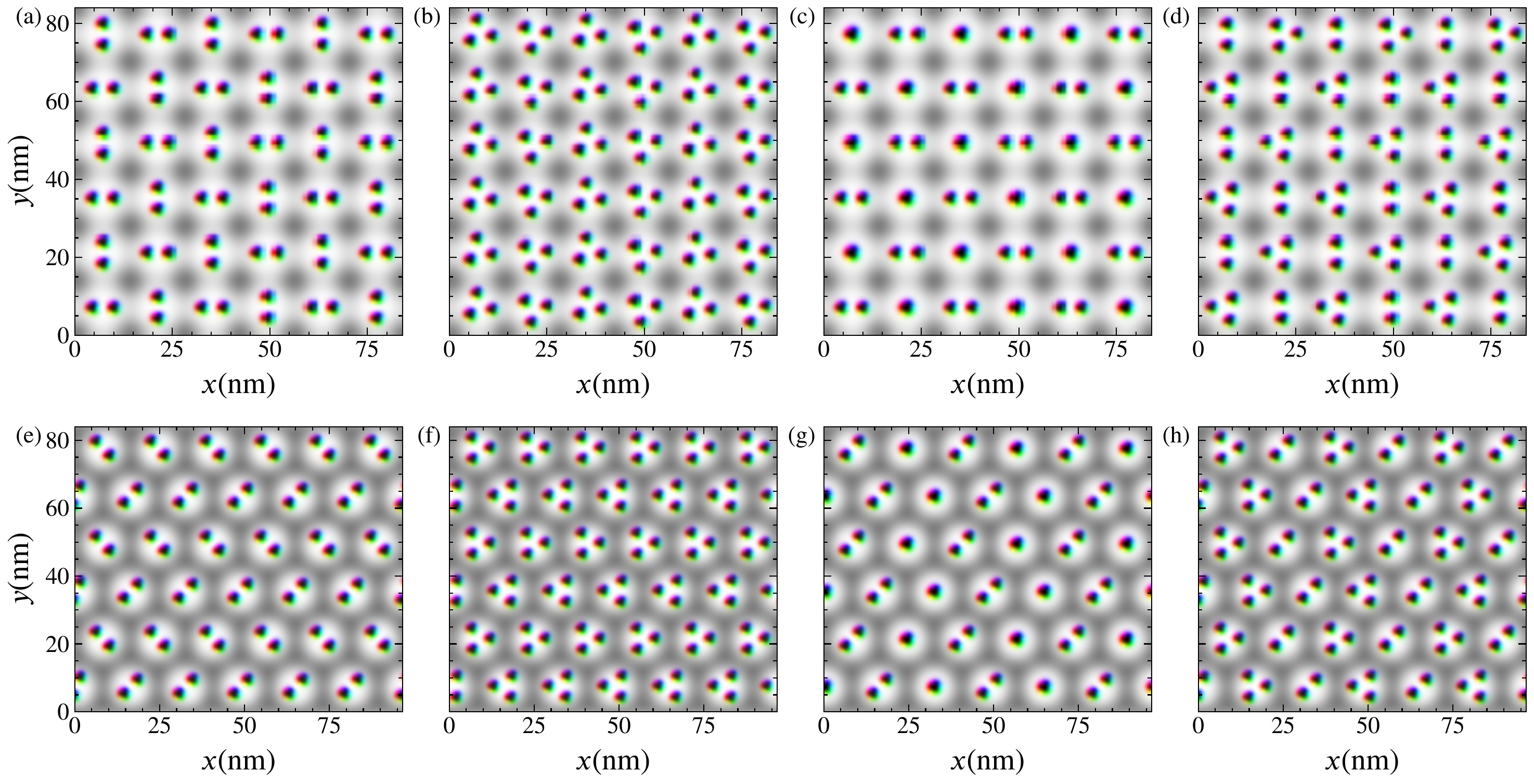}
  \caption{Different skyrmion arrangements for skyrmions on
    (a-d) square and (e-h) triangular
    substrates from an atomistic model.
    The sample perpendicular magnetic
    anisotropy (PMA) is represented by a gray overlay, where darker
    areas have higher PMA and brighter areas have lower PMA.
    (a) Antiferromagnetic and (e) herringbone
    dimer ordering at $N_\mathrm{sk}/N_m=2$.
    (b, f) Alternating column trimer arrangement
    at $N_\mathrm{sk}/N_m=3$. (c, g) Bipartite
    monomer-dimer arrangement with ferromagnetically aligned dimers
    at $N_\mathrm{sk}/N_m=3/2$.
    (d, h) Bipartite dimer-trimer 5/2 arrangement at
    $N_\mathrm{sk}/N_m=5/2$.}
  \label{fig:1}
\end{figure}

In this work we consider two anisotropy patterns, a square substrate
given by
\begin{equation}\label{eq:3}
  K_S(x, y)=\frac{K_0}{4}\left[\cos\left(\frac{2\pi
      x}{a_0}\right)+\cos\left(\frac{2\pi y}{a_0}\right) + 2\right]
  \ ,
\end{equation}
and a triangular substrate given by
\begin{equation}\label{eq:4}
  K_T(x, y)=\frac{2K_0}{9}\left[3+\sum_{i=1}^{3}\cos\left(\frac{2\pi
      b_i}{a_0}\right)\right] \ ,
\end{equation}
where $K_0$ is the anisotropy depth and $a_0=14$ nm is the substrate
lattice constant for both patterns. In the triangular pattern,
$b_i=x\cos\theta_i-y\sin\theta_i+a_0/2$ with $\theta_1=\pi/6$,
$\theta_2=\pi/2$ and $\theta_3=5\pi/6$. Both patterns produce
anisotropies falling in the range $0\leq K(x, y)\leq K_0$,
and both the square and
triangular substrates have a total of
$N_m=36$ anisotropy minima. For
the triangular substrate, we change
the sample dimensions to 96.5 nm $\times$ 84 nm in
order to properly apply the periodic boundary conditions.

The skyrmion velocity is computed using the emergent electromagnetic
fields \cite{seki_skyrmions_2016,schulz_emergent_2012}:

\begin{eqnarray}\label{eq:5}
  E^\mathrm{em}_i=\frac{\hbar}{e}\mathbf{m}\cdot\left(\frac{\partial
    \mathbf{m}}{\partial i}\times\frac{\partial \mathbf{m}}{\partial
    t}\right)\\ B^\mathrm{em}_i=\frac{\hbar}{2e}\varepsilon_{ijk}\mathbf{m}\cdot\left(\frac{\partial
    \mathbf{m}}{\partial j}\times\frac{\partial \mathbf{m}}{\partial
    k}\right) \ ,
\end{eqnarray}
where $\varepsilon_{ijk}$ is the totally anti-symmetric tensor. The
skyrmion drift velocity, $\mathbf{v}_d$, is then calculated using
$\mathbf{E}^\mathrm{em}=-\mathbf{v}_d\times\mathbf{B}^\mathrm{em}$. The
skyrmion Hall angle is computed using
$\theta=\arctan\left(\left\langle v_y\right\rangle/\left\langle v_x\right\rangle\right)$

We fix the following values in our simulations:
$\mu\mathbf{H}=0.6(D^2/J)(-\mathbf{\hat{z}})$, $\alpha=0.3$, and
$p=-1.0$. The material parameters are $J=1$ meV, $D=0.5J$, and
$K_0=0.2J$ unless otherwise mentioned. For each simulation, the
system is initialized in one of the configurations illustrated in
Fig.~\ref{fig:1}. The numerical integration of Eq.~\ref{eq:2} is
performed using a fourth order Runge-Kutta method over 200 ns.

In our system, we consider STTs that create a force perpendicular to
the current direction. Thus, just at depinning,
when the skyrmions travel in the direction of
the driving force they have a Hall angle of $90^\circ$,
which decreases to the intrinsic value at higher drives.
When we perform simulations on a clean sample, we find that the
skyrmions move at $\theta_\mathrm{sk}^\mathrm{int}=26^\circ$ with
respect to the current, which corresponds to an angle of $\theta=64^\circ$
with respect to the driving force.

\subsection{Particle based simulations}

We also simulate the particle based model developed by Lin {\it et
  al.}~\cite{lin_particle_2013}. The equation of motion governing the
dynamics is given by
\begin{equation}\label{eq:6}
  \alpha_d\mathbf{v}_i+\alpha_m\hat{\mathbf{z}}\times\mathbf{v}_i=\sum_{i\neq
    j}\mathbf{F}_\mathrm{sk}(\mathbf{r}_{ij})+\mathbf{F}_\mathrm{P}(\mathbf{r}_i)+\hat{\mathbf{z}}\times\mathbf{F}_D
  \ ,
\end{equation}
where $\mathbf{v}_i$ is the velocity of skyrmion $i$. The first term on
the left is the damping term of strength $\alpha_d$. The second term is the
Magnus force of strength $\alpha_m$. The intrinsic
Hall angle is given by
$\theta_\mathrm{sk}^\mathrm{int}=\arctan(\alpha_m/\alpha_d)$. In
order to match the particle based simulations
with our atomistic based simulations,
we performed atomistic simulations on clean samples and
obtained an intrinsic Hall angle of
$\theta_\mathrm{sk}^\mathrm{int}=26^\circ$; this value was then used
to determine the values of $\alpha_m=\sin(64^\circ)$ and
$\alpha_d=\cos(64^\circ)$ based on the fact that the skyrmions
move with an angle of $64^\circ$ with respect to the driving force.
The first term on the right side of
Eq.~\ref{eq:6} is the skyrmion-skyrmion interaction, given by
$\mathbf{F}_\mathrm{sk}(\mathbf{r}_{ij})=-U_\mathrm{sk}K_1(r_{ij})\hat{\mathbf{r}}_{ij}$,
where $U_\mathrm{sk}=1$ is the interaction strength and $K_1$ is the
modified first order Bessel function. The second term is the
interaction with the underlying substrate potential, represented
by the same equations,
Eqs.~\ref{eq:3} and \ref{eq:4}, as in the atomistic model,
with the adjustment $K_0\to A$ where $A=0.75$. By taking
$a_0=6$ we obtain substrates with $N_m=36$ potential minima. The
last term is the interaction with an external drive,
$\mathbf{F}_D=F_D\hat{\mathbf{x}}$, corresponding
to the action of an STT
current on magnetic skyrmions
\cite{iwasaki_universal_2013,everschor-sitte_perspective_2018,feilhauer_controlled_2020}.
Our simulation box is of size $36\times36$ with periodic boundary
conditions along the $x$ and $y$ directions. The sample size is adjusted
to $36(2/\sqrt{3})\times36$ for the triangular substrate in order to
properly apply the periodic boundary conditions.

\begin{figure}
  \centering
  \includegraphics[width=0.7\columnwidth]{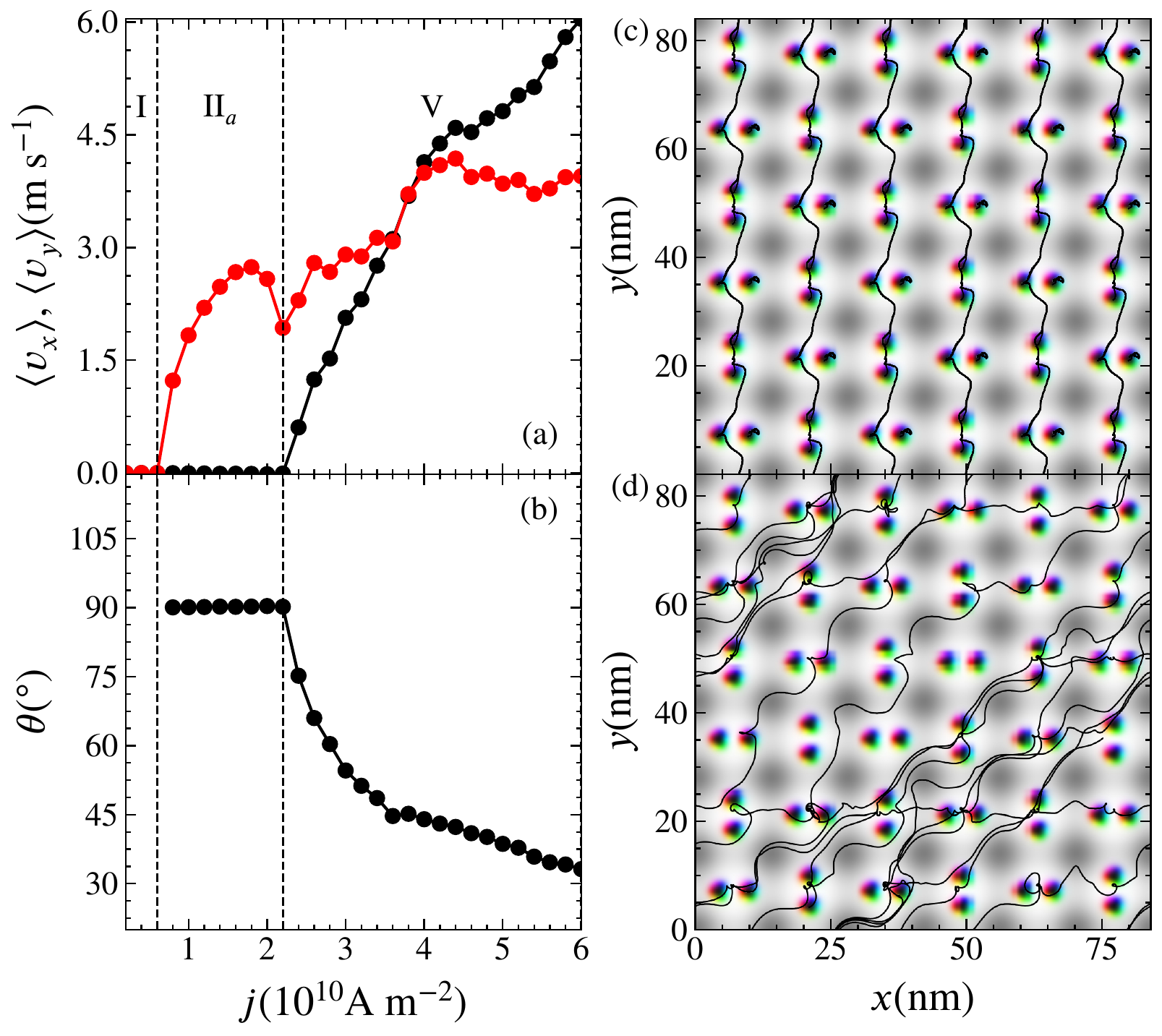}
  \caption{
    (a) Average skyrmion velocities $\left\langle v_x\right\rangle$ (black)
    and $\left\langle v_y\right\rangle$ (red) and
    (b) the corresponding skyrmion Hall angle $\theta$
    vs applied current $j$ from an atomistic model
    for the dimer $N_\mathrm{sk}/N_m = 2.0$ filling on a square substrate
    shown in Fig.~\ref{fig:1}(a).
    I: pinned phase. II$_a$: 90$^\circ$ transverse motion.
    V: partially disordered motion.
    (c, d) Skyrmion
    trajectories at representative drives for the
    same system. (c) Trajectories
    of all skyrmions in phase II$_a$
    at $j=1.4\times10^{10}$~A m$^{-2}$. (d)
    Trajectory of a single skyrmion in phase V at
    $j=5\times10^{10}$~A m$^{-2}$.}
  \label{fig:2}
\end{figure}

\section{Dynamics of Skyrmion Molecular Crystals on Square Substrates}

We first consider atomistic simulations of skyrmions on a square
substrate, illustrated in Fig.~\ref{fig:1}(a-d).
At $N_\mathrm{sk}/N_m = 2$ in Fig.~\ref{fig:1}(a),
the skyrmions form
dimers with the dimer center located at the trap minimum.
Adjacent dimers
are perpendicular to each other forming what we describe as
antiferromagnetic ordering, which is similar to the ordering
observed for dimerized colloidal particles
on a square substrate
\cite{reichhardt_novel_2002,agra_theory_2004}.
The colloidal particles interact with a
short-range repulsion, and in that system the
antiferromagnetic arrangement arises due to minimization of
an effective quadrupole moment of the adjacent dimers
\cite{agra_theory_2004}.
Skyrmions can
also be treated as repulsive particles, so they form a structure
similar to that found for the colloidal system.
At $N_\mathrm{sk}/N_m = 3$ in
Fig.~\ref{fig:1}(b), trimers form with an orientation that
alternates from one column to the next.
Figure~\ref{fig:1}(c) shows a bipartite state at
$N_\mathrm{sk}/N_m = 3/2$, where an alternating mixture of
monomers and dimers is present. Here,
all of the dimers are aligned and exhibit
ferromagnetic ordering.
Similarly,
at $N_\mathrm{sk}/N_m = 5/2$ in Fig.~\ref{fig:1}(d),
in an alternating arrangement of dimers and trimers,
the dimers are all aligned with each other and the trimers are also
aligned with a slight tilt.
Previous studies of skyrmion molecular
crystals were performed for skyrmions coupled to a triangular substrate.
We observe
similar orderings as shown in Figs.~\ref{fig:1}(e-h).
At $N_\mathrm{sk}/N_m = 2$ and $N_\mathrm{sk}/N_m = 3$, the tilt angles of
the dimers and trimers in Fig.~\ref{fig:1}(e) and (f), respectively,
alternate from site to site, while
the dimers in both the
$N_\mathrm{sk}/N_m= 3/2$ and $N_\mathrm{sk}/N_m = 5/2$ states
in Fig.~\ref{fig:1}(g) and (h) have
ferromagnetic or aligned ordering.
As we vary the filling, we find many other states beyond those illustrated
in Fig.~\ref{fig:1}, which we will describe in a future work.

\begin{figure}
  \centering
  \includegraphics[width=\columnwidth]{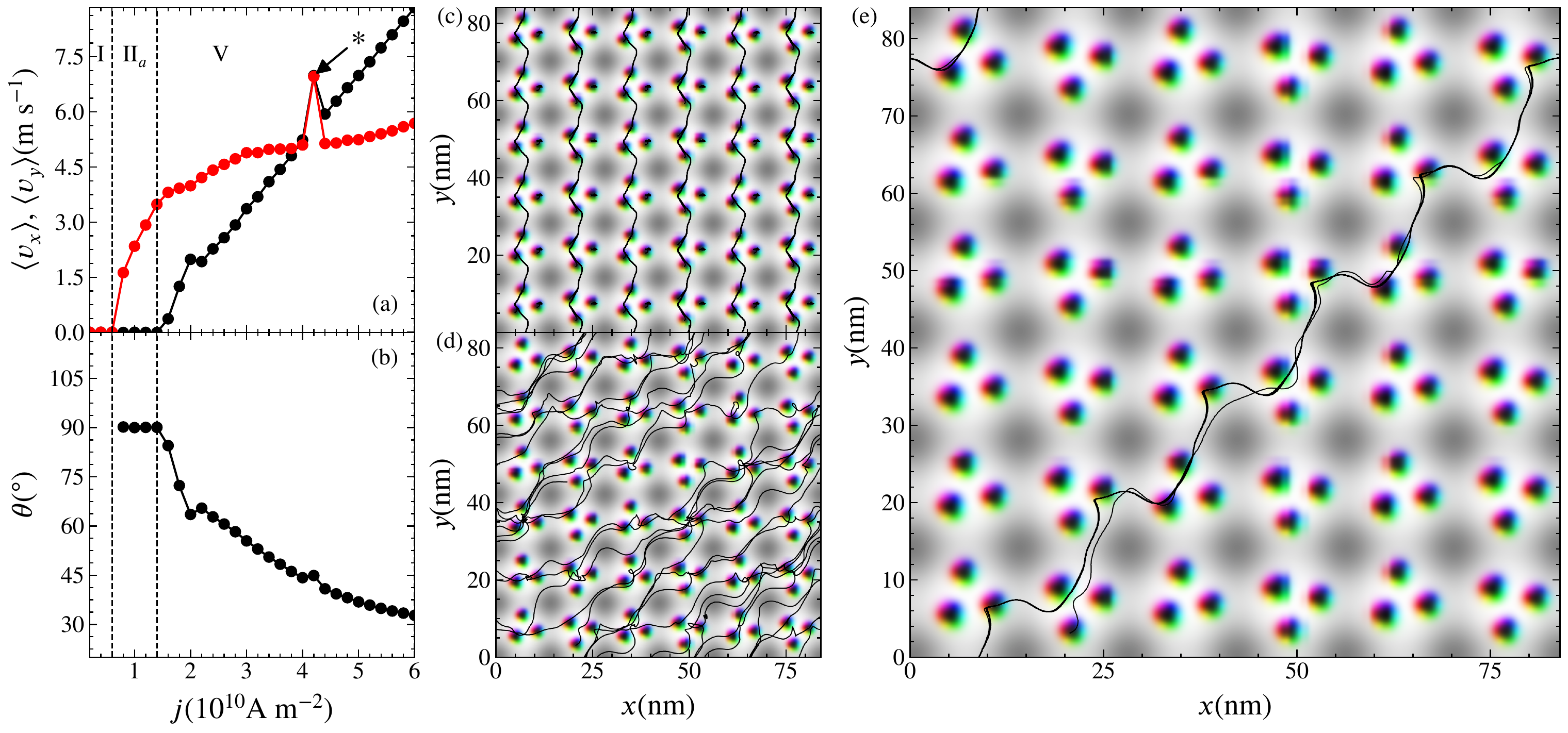}
  \caption{(a) Average skyrmion velocities
    $\left\langle v_x\right\rangle$ (black) and $\left\langle v_y\right\rangle$
    (red) and (b) the corresponding skyrmion Hall angle $\theta$
    vs $j$ from an atomistic model
    for the trimer $N_\mathrm{sk}/N_m = 3.0$ filling on a square
    substrate shown in
    Fig.~\ref{fig:1}(b).
    I: pinned phase. II$_a$: 90$^\circ$ transverse motion.
    V: partially disordered motion.
    The $*$ symbol indicates the point
    at $j=4.2\times10^{10}$~A m$^{-2}$ where the velocities
    peak and the motion is directionally locked along $45^\circ$.
    (c, d, e) Skyrmion trajectories at representative drives for the
    same system.
    (c) Trajectories of all skyrmions in phase II$_a$ at
    $j=1\times10^{10}$~A m$^{-2}$. (d) Trajectory of a single
    skyrmion in phase V at $j=5\times10^{10}$~A m$^{-2}$.
    (e) Trajectory of a single skyrmion in phase
    V at $j=4.2\times10^{10}$~A m$^{-2}$, the directionally
    locked current marked
    by a $*$ in panel (a).
  }
  \label{fig:3}
\end{figure}

In Fig.~\ref{fig:2}(a) we plot the average skyrmion velocities
$\left\langle v_x\right\rangle$ and $\left\langle v_y\right\rangle$
versus current $j$ for the $N_\mathrm{sk}/N_m = 2$
dimer system from Fig.~\ref{fig:1}(a),
and in Fig.~\ref{fig:2}(b) we show the
corresponding Hall angle $\theta$ versus $j$. In
the absence of a substrate, the skyrmions would flow
at $\theta = 26^\circ$.
We find that for
$j \leq 0.6\times10^{10}$ A m $^{-2}$,
the skyrmions are pinned, which we label region I in Fig.~\ref{fig:2}(a).
The initial depinning occurs into a state
where $\left\langle v_y\right\rangle$ increases with increasing $j$
but $\left\langle v_x\right\rangle = 0.0$,
indicating that the flow is perpendicular to the
current.
This transverse motion appears when $0.6\times10^{10}~\mathrm{A m}^{-2}< j\leq 2.2\times 10^{10}$ A m$^{-2}$,
which we label region II$_a$.
As $j$ increases further,
$\langle v_x\rangle$ begins to increase
and $\theta$ decreases from $90^\circ$ toward the
intrinsic Hall angle value of 26$^\circ$. We label this region V flow.
There is small window in which $\left\langle v_x\right\rangle$ and
$\left\langle v_y\right\rangle$ are locked
together, corresponding to directional locking
along $45^\circ$, the major symmetry direction
of the square substrate.
Just below the transition from region
II$_a$ to region V,
$\left\langle v_y\right\rangle$
decreases with increasing $j$, indicating
negative differential mobility.
A similar velocity dip has
been observed at dynamic transitions for vortex flow in superconductors
with periodic pinning \cite{Reichhardt17}, as
well as for a single
skyrmion moving over a periodic substrate at transitions into and out
of certain symmetry locking directions \cite{Reichhardt15a}.

\begin{figure}
  \centering
  \includegraphics[width=0.7\columnwidth]{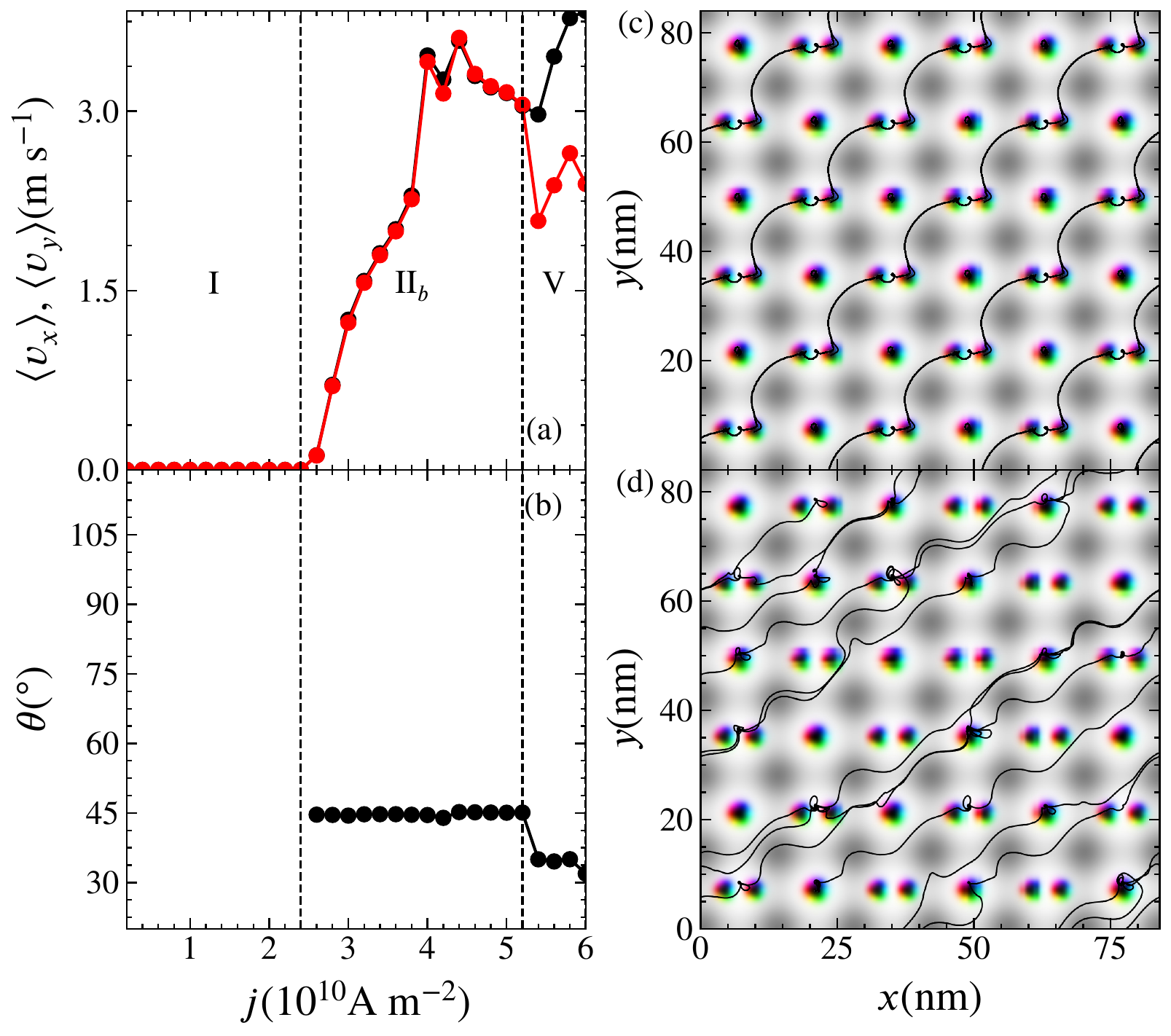}
  \caption{Average skyrmion velocities
    $\left\langle v_x\right\rangle$ (black) and $\left\langle v_y\right\rangle$
    (red) and (b) the corresponding skyrmion Hall angle $\theta$ vs
    $j$ from an atomistic model
    for the monomer-dimer lattice at $N_\mathrm{sk}/N_m = 3/2$ on a
    square substrate
    from Fig.~\ref{fig:1}(c).
    I: pinned phase. II$_b$: 45$^\circ$ directionally locked motion.
    V: partially disordered motion with no locking.
    (c, d) Skyrmion trajectories at representative drives
    for the same system. (c) Trajectories of all skyrmions
    in phase II$_b$ at $j=3\times10^{10}$~A m$^{-2}$.
    (d) Trajectory of a single skyrmion in phase V at
    $j=6\times10^{10}$~A m$^{-2}$.
  }
  \label{fig:4}
\end{figure}

In Fig.~\ref{fig:2}(c) we illustrate the skyrmion locations and
trajectories in phase II$_a$ at $j=1.4\times10^{10}$~A m$^{-2}$, where
flow occurs via soliton motion in which an individual skyrmion jumps from
one well to the next and triggers the jump of another single skyrmion.
When the skyrmion jumps into a well with a dimer that is aligned along
the $x$ direction, the dimer rotates into the $y$ direction and the upper
skyrmion jumps out of the well. If the dimer is already aligned in the $y$
direction before the arrival of the jumping skyrmion, the upper skyrmion
jumps into the next well and no rotation occurs.
At higher drives, the system enters phase V where the
flow is more disordered and occurs along both the $x$ and $y$
directions.
The phase V motion is shown in
Fig.~\ref{fig:2}(d) at $j=5\times10^{10}$~A m$^{-2}$, where we draw
only one skyrmion trajectory for clarity. Temporary locking motion
along $45^\circ$ is intermingled with more disordered flow,
and some of the dimers remain pinned.
As the drive increases,
more of the dimers participate in the flow, which
gradually rotates into a direction
closer to that of the intrinsic Hall angle.
In a sample just off commensuration, where a few skyrmions are removed
or added, we find
a similar trend in which the angle of skyrmion motion is
initially large and then decreases as the drive increases;
however, the presence of vacancies or interstitials
produces much stronger directional locking of the flow compared
to the commensurate case \cite{Souza24}.

\begin{figure}
  \centering
  \includegraphics[width=0.7\columnwidth]{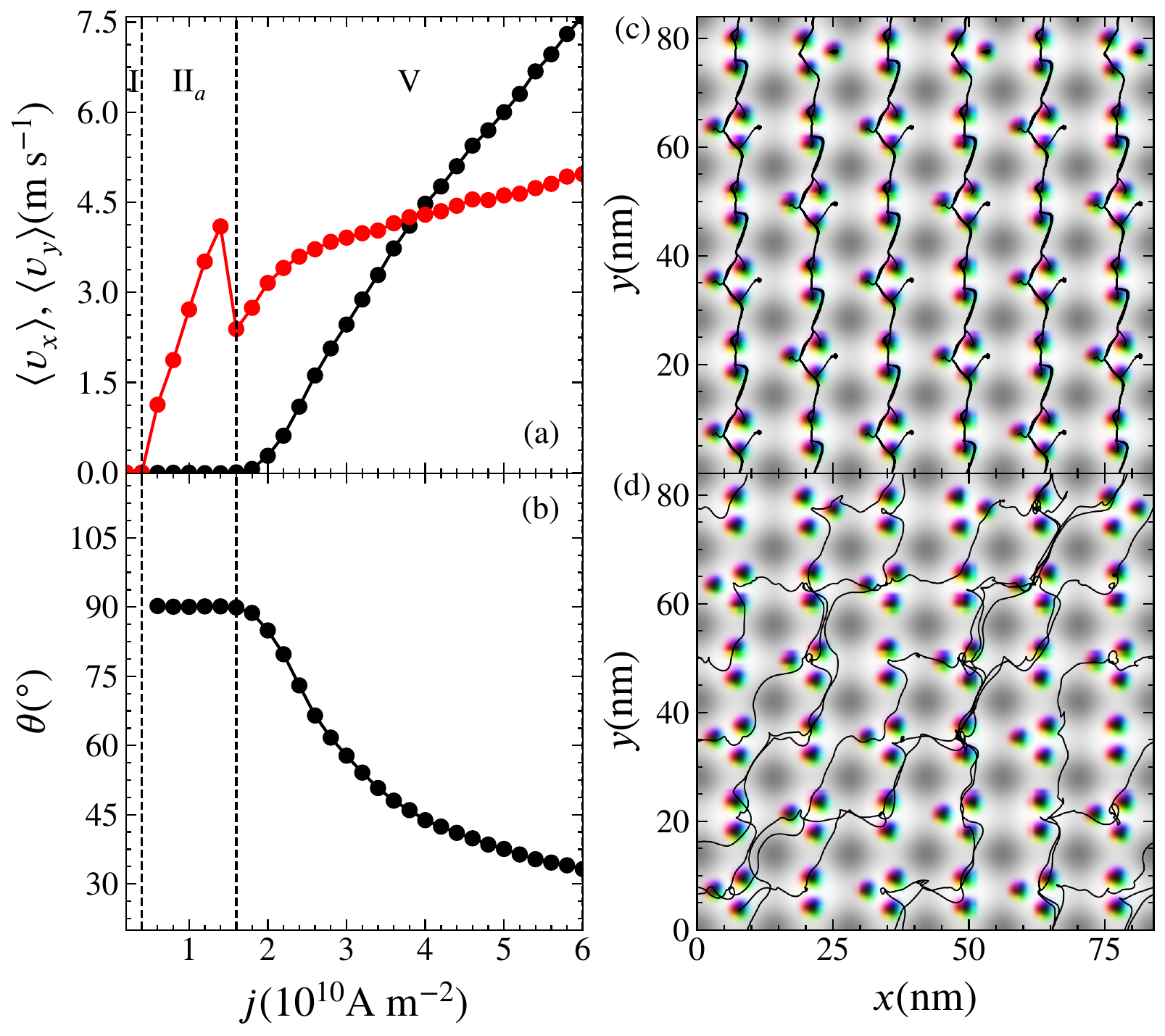}
  \caption{Average skyrmion velocities
    $\left\langle v_x\right\rangle$ (black) and $\left\langle v_y\right\rangle$
    (red) and (b) the corresponding skyrmion Hall angle $\theta$
    vs $j$ from an atomistic model
    for the dimer-trimer $N_\mathrm{sk}/N_m = 5/2$ filling on
    a square substrate from
    Fig.~\ref{fig:1}(d).
    I: pinned phase. II$_a$: 90$^\circ$ transverse motion.
    V: partially disordered motion.
    (c, d) Skyrmion trajectories at representative drives for the same
    system. (c) Trajectories of all skyrmions
    in phase II$_a$ at $j=1\times10^{10}$~A m$^{-2}$.
    (d) Trajectory of a single skyrmion in phase V at
    $j=3\times10^{10}$~A m$^{-2}$.
  }
  \label{fig:5}
\end{figure}

Figure~\ref{fig:3}(a,b) shows the average velocities
$\left\langle v_x\right\rangle$ and $\left\langle v_y\right\rangle$ along
with $\theta$ versus applied current $j$
for the $N_\mathrm{sk}/N_m=3$ trimer lattice system
from Fig.~\ref{fig:1}(b).
The pinned phase I appears
for $j \leq 0.6\times 10^{10}$ A m $^{-2}$, while
the transverse motion phase II$_a$
spans the range
$0.6\times 10^{10}~\mathrm{A m}^{-2} < j \leq 1.4 \times 10^{10}$ A m $^{-2}$.
At higher drives we find phase V motion.
There is also a small region around
$j=4.2\times10^{10}$~A m$^{-2}$ where both velocities peak and
strong 45$^\circ$ directional
locking is present.
In Fig.~\ref{fig:3}(c) we illustrate the skyrmion
motion in phase II$_a$ at
$j=1\times10^{10}$~A m$^{-2}$.
The flow is in the form of a soliton,
where a single skyrmion jumps from one well to the
next and causes another skyrmion to jump to the next well.
One skyrmion in each trimer does not participate in the motion,
as indicated by
the absence of trajectories for the skyrmions on the right side of
each well.
Figure~\ref{fig:3}(d)
shows the motion of a single skyrmion during disordered phase V
flow at
$j=5\times10^{10}$~A m$^{-2}$, where some pinned skyrmions are
still present.
The $45^\circ$ directional locking
state at $j=4.2\times10^{10}$ A m$^{-2}$ is plotted in
Fig.~\ref{fig:3}(e), where only a single skyrmion trajectory
is drawn.
Instead of soliton flow,
all of the skyrmions are flowing in this state.
For higher drives, the flow becomes more disordered and
some of the skyrmions repin, resulting in
a velocity drop above the directionally locked state.

In Fig.~\ref{fig:4}(a, b) we show $\left\langle v_x\right\rangle$,
$\left\langle v_y\right\rangle$, and $\theta_\mathrm{sk}$ vs $j$ for
the system in Fig.~\ref{fig:1}(c) at the $N_\mathrm{sk}/N_m = 3/2$
filling. In this case, there is a much larger pinned phase I
that persists up to
$j = 2.4\times10^{10}$ A m $^{-2}$, followed by extended regions of flow
locked along $45^\circ$
for $ 2.4\times10^{10}~\mathrm{A m}^{-2} < j \leq 5.2\times10^{10}$ A m $^{-2}$,
which we label phase II$_b$.
At higher drives, the system transitions to disordered phase V flow,
and a velocity drop appears at the II$_b$ to V transition.
There is a well-defined step feature
in $\theta$
associated with the $45^\circ$ locking state.
In
Fig.~\ref{fig:4}(c), we show the skyrmion trajectories for the
phase II$_b$ flow at $j=3\times10^{10}$~A m$^{-2}$.
A solitonic flow passes only through the dimer wells and the motion
runs along $45^\circ$.
The sites containing only one skyrmion remain permanently pinned.
Figure~\ref{fig:4}(d) shows the trajectory of a single skyrmion
at $j=6\times10^{10}$~A m$^{-2}$ in phase V, where the
motion is more disordered and there is a mixture of pinned monomers and dimers.
Compared to the $N_\mathrm{sk}/N_m = 2.0$ system,
the $N_\mathrm{sk}/N_m = 3/2$ does not have a regime in which the motion
runs completely transverse to the applied current. Instead, the transverse
flow state has been replaced by a pinned phase.

\begin{figure}
  \centering
  \includegraphics[width=0.7\columnwidth]{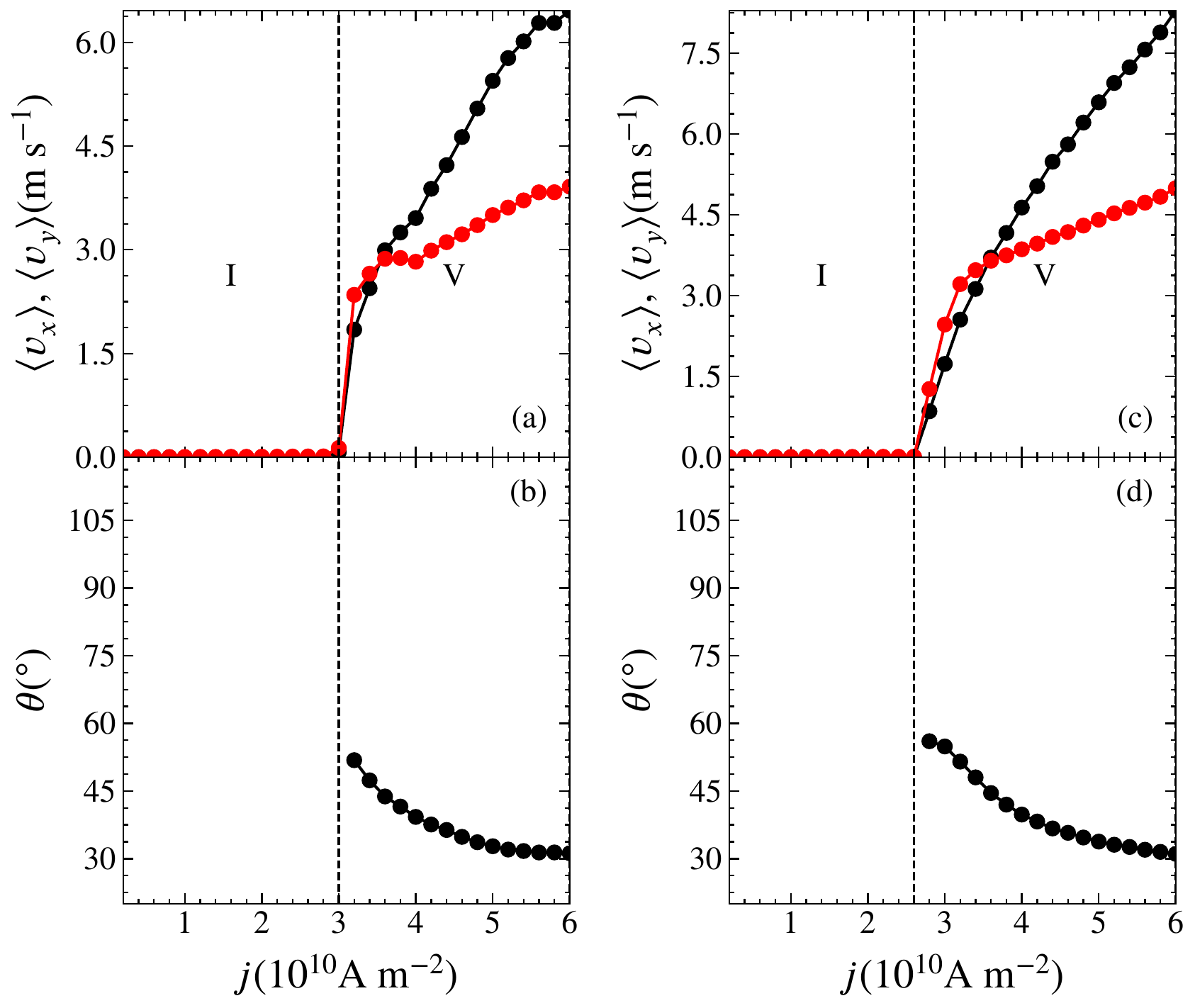}
  \caption{
    (a, c) Average skyrmion velocities
    $\left\langle v_x\right\rangle$ (black) and
    $\left\langle v_y\right\rangle$ (red) and
    (b, d) the corresponding skyrmion Hall angle $\theta$ vs $j$
    from an atomistic model on a triangular substrate.
    (a, b) The $N_\mathrm{sk}/N_m = 2.0$ system from Fig.~\ref{fig:1}(e).
    (c, d) The $N_\mathrm{sk}/N_m = 3.0$ system from Fig.~\ref{fig:1}(f).
    I: pinned phase.
    V: disordered motion.
  }
  \label{fig:6}
\end{figure}

For the dimer-trimer system in Fig.~\ref{fig:1}(d) at
a filling of $N_\mathrm{sk}/N_m = 5/2$,
in Fig.~\ref{fig:5}(a, b) we plot
$\left\langle v_x\right\rangle$, $\left\langle v_y\right\rangle$, and
$\theta$ versus $j$.
The pinned region is reduced in size and appears when
$j \leq 0.4\times10^{10}$~A m$^{-2}$,
while there is a region of transverse II$_a$ flow
over the range
$0.4\times10^{10}~\mathrm{A m}^{-2} < j \leq 1.6\times10^{10}$~A m$^{-2}$.
As the drive increases,
$\left\langle v_y\right\rangle$ increases linearly
throughout region II$_a$
and then passes through
a sharp drop when the system enters the disordered flow phase V.
For this filling, we do not observe any
$45^\circ$ directional locking, and
$\theta$ smoothly approaches the intrinsic value as the current increases.
In Fig.~\ref{fig:5}(c), skyrmion trajectories from
the II$_a$ phase at $j=1\times10^{10}$~A m$^{-2}$
indicate that the motion consists of $y$ direction soliton
flow.
The trajectory of a single skyrmion in the disordered phase V flow
state at
$j=3\times10^{10}$~A m$^{-2}$ appears in
Fig.~\ref{fig:5}(d).

\section{Dynamics of Skyrmion Molecular Crystals on Triangular Substrates}

We next consider the case of driven skyrmion molecular crystals on
triangular substrates. We use the same parameters as the square
substrate; the only difference is the sample size, which is
changed to $L_y = 84$ nm and $L_x = 96.5$ nm.
In general, we find a strong pinning
effect for triangular substrates. The flow is more disordered,
and the $90^\circ$ locking flow phases that appear
on the square substrate do not occur for the triangular substrate.

\begin{figure}
  \centering
  \includegraphics[width=\columnwidth]{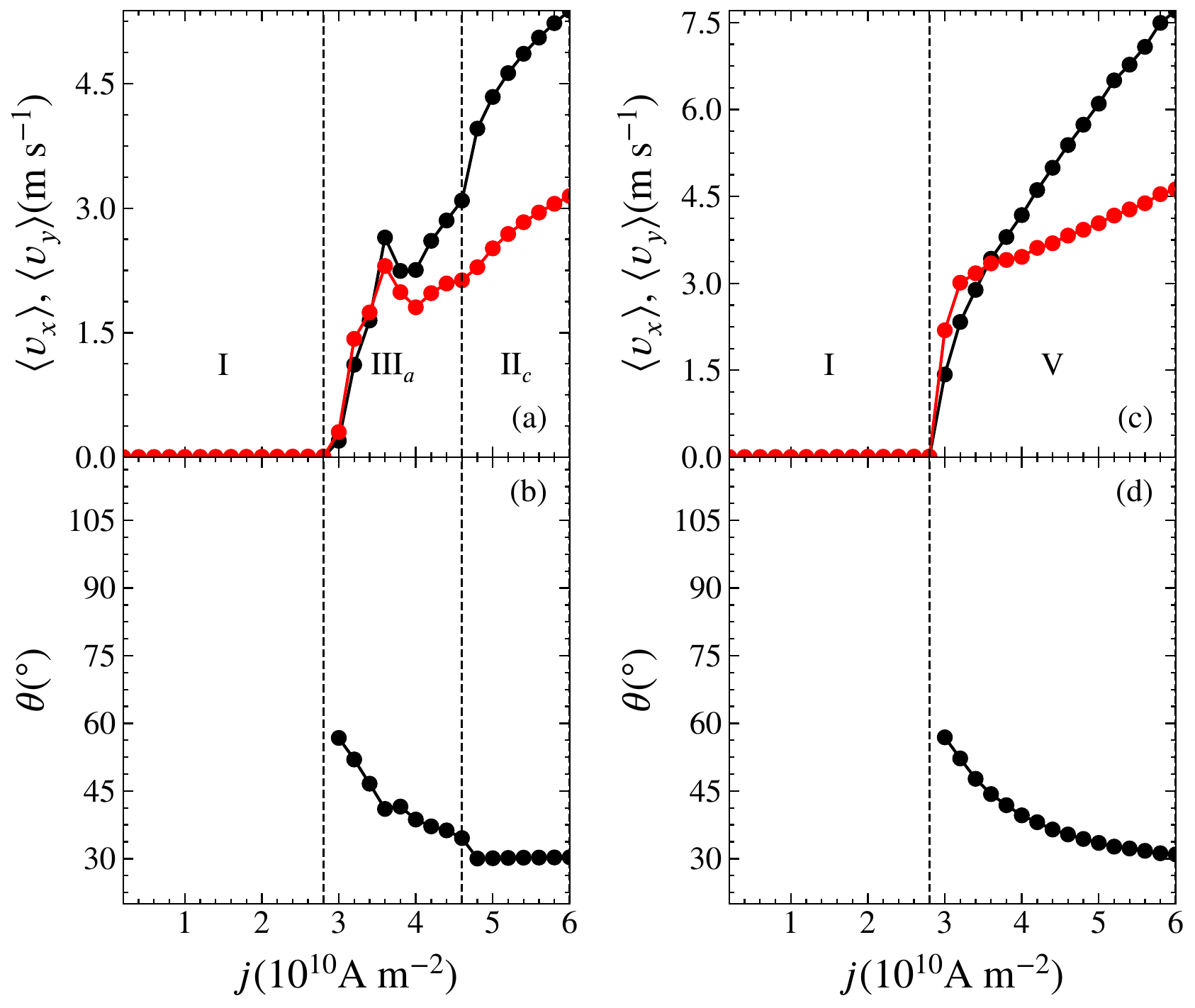}
  \caption{Skyrmion trajectories at representative drives
    for a
    triangular substrate with the same parameters as the square
    substrate in Fig.~\ref{fig:1}(a-d) from an atomistic model.
    (a) Trajectory of a single skyrmion for the dimer state
    with $N_\mathrm{sk}/N_m=2.0$ at
    $j=3.2\times10^{10}$ A m$^{-2}$.
    (b) Trajectory of a single skyrmion for the trimer state
    with $N_\mathrm{sk}/N_m=3.0$ at
    $j=3.2\times10^{10}$ A m$^{-2}$.
    (c) Trajectory of a single skyrmion for the dimer state
    with $N_\mathrm{sk}/N_m=2.0$ at
    $j=5\times10^{10}$ A m$^{-2}$.
    (d) Trajectory of a single skyrmion for the trimer state
    with $N_\mathrm{sk}/N_m=3.0$ at
    $j=5\times10^{10}$ A m$^{-2}$.
    (e) Trajectories of all skyrmions
    for the $N_\mathrm{sk}/N_m = 3/2$ monomer-dimer filling
    in the directional locking phase II$_c$ at
    $j=5\times10^{10}$ A m$^{-2}$
    where
    the motion is along $30^\circ$.
    (f) Trajectory of a single skyrmion for the
    $N_\mathrm{sk}/N_m = 5/2$ dimer-trimer filling
    in the disordered flow state at $j=5\times10^{10}$ A m$^{-2}$.
  }
  \label{fig:8}
\end{figure}

In Fig.~\ref{fig:6}(a, b) we plot $\left\langle v_x\right\rangle$,
$\left\langle v_y\right\rangle$, and $\theta$
versus $j$ for the
$N_\mathrm{sk}/N_m = 2.0$ dimer filling for skyrmions on a triangular
substrate that appears in Fig.~\ref{fig:1}(e).
There is an extended pinned regime that ends at
$j = 3\times10^{10}$ A m $^{-2}$
and is followed by a disordered flow state in which
both $\langle v_x\rangle$ and $\langle v_y\rangle$ increase with
increasing drive.
Just above depinning,
$\theta$ takes a value slightly lower than
$60^\circ$,
corresponding to a partial locking of the skyrmion motion to
the $60^\circ$ symmetry direction of the triangular substrate.
In
Fig.~\ref{fig:8}(a) we show the trajectories of a single skyrmion
at $j=3.2\times10^{10}$ A m$^{-2}$ where the flow is partially disordered
and there is partial channeling along
$60^\circ$.
At $j=5\times10^{10}$ A m$^{-2}$, the plot of a single
skyrmion trajectory in
Fig.~\ref{fig:8}(c) indicates that
the motion is much more disordered compared to the phase V flow in the
square substrate illustrated in \ref{fig:2}(e),
and there is some tendency for the skyrmions to move along $60^\circ$.
As $j$ increases, Fig.~\ref{fig:6}(b) shows that
$\theta$ gradually
decreases toward the intrinsic value.

We show
$\left\langle v_x\right\rangle$, $\left\langle v_y\right\rangle$,
and $\theta$ versus $j$ for
the $N_\mathrm{sk}/N_m = 3.0$ trimer filling of a
triangular substrate in Fig.~\ref{fig:6}(c, d).
Similar to what we observed for the dimer state,
there is no transverse motion, and the flow just above depinning initially
runs along a direction close to a major substrate symmetry direction.
The Hall angle
is close to $60^\circ$ above depinning and decreases toward
the intrinsic value as the drive increases.
In Fig.~\ref{fig:8}(b) we plot the
trajectory of a single skyrmion at
$j=3.2\times10^{10}$ A m$^{-2}$ where the flow is partially
disordered and partial channeling along $60^\circ$ is present.
A single trajectory at a higher drive of
$j=5\times10^{10}$ A m$^{-2}$, shown in
Fig.~\ref{fig:8}(d), indicates that the
channeling is increasingly lost and
the flow is becoming more disordered.

\begin{figure}
  \centering
  \includegraphics[width=0.7\columnwidth]{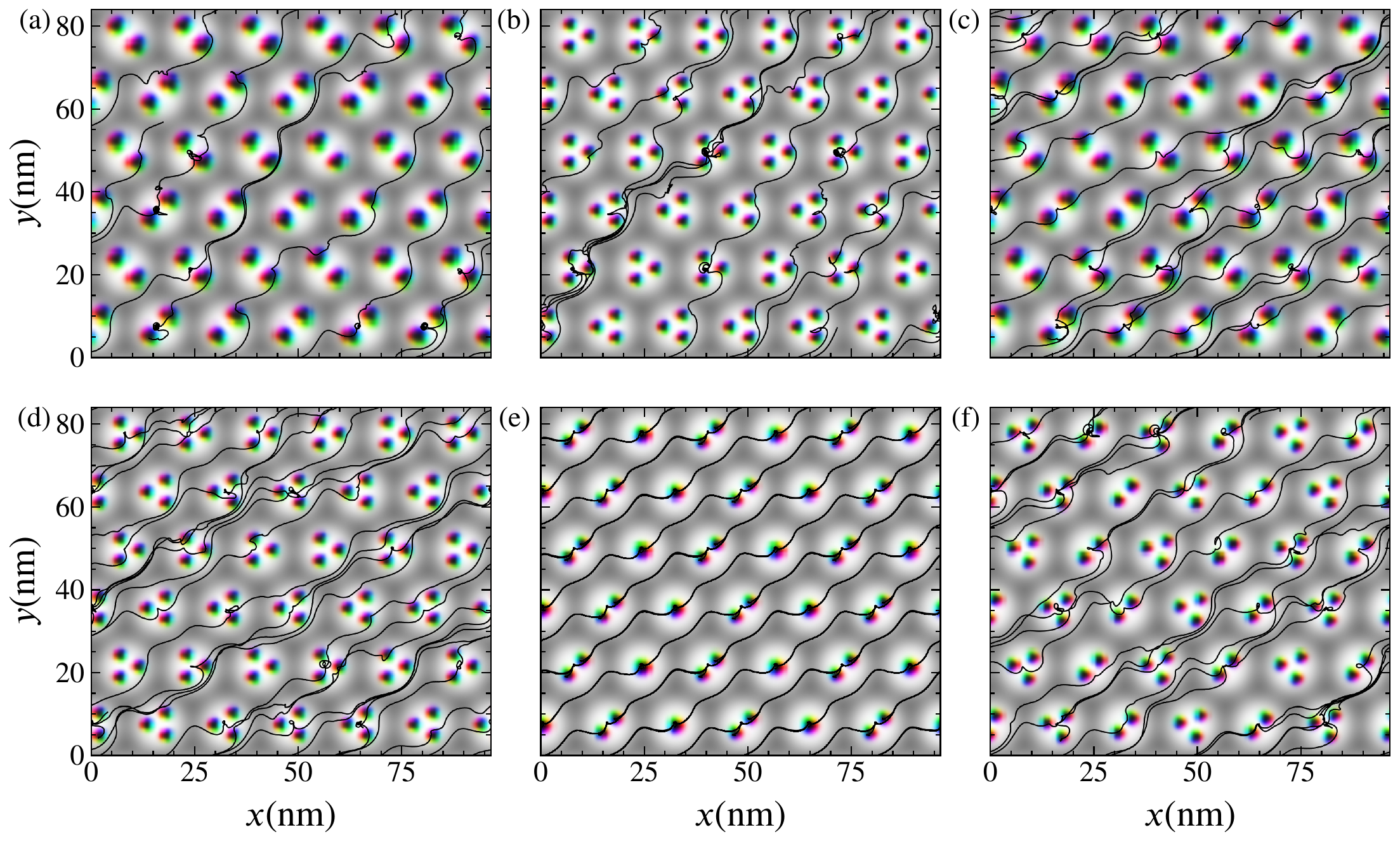}
  \caption{(a, c) Average skyrmion velocities
    $\left\langle v_x\right\rangle$ (black) and
    $\left\langle v_y\right\rangle$ (red) and
    (b, d) the corresponding skyrmion Hall angle
    $\theta$ vs $j$ from an atomistic model on a triangular substrate.
    (a, b) The $N_\mathrm{sk}/N_m = 3/2$ monomer-dimer filling from
    Fig.~\ref{fig:1}(g).
    (c, d) The $N_\mathrm{sk}/N_m = 5/2$ dimer-trimer filling
    from Fig.~\ref{fig:1}(h).
    I: pinned phase.
    II$_c$: 30$^\circ$ directional locking state.
    III$_a$: transitional soliton flow regime with no directional locking.
    V: disordered motion.
    }
  \label{fig:7}
\end{figure}

Figure~\ref{fig:7}(a, b) shows $\left\langle v_x\right\rangle$,
$\left\langle v_y\right\rangle$, and $\theta$ versus $j$
for the $N_\mathrm{sk}/N_m = 3/2$ monomer-dimer filling on the
triangular substrate.
For $j > 4.6 \times 10^{10}$ A m $^{-2}$, the flow locks
along $30^\circ$ and creates a step feature in $\theta$. We label this
directionally locked state phase II$_c$.
Between depinning and the onset of phase II$_c$ is a transitional
phase we term
III$_a$ in which soliton-like motion occurs but there is no
directional locking.
In Fig.~\ref{fig:8}(e), the plot of all skyrmion
trajectories at $j = 5\times10^{10}$ A m $^{-2}$
demonstrates that
the II$_c$ ordered motion is a directionally locked phase
with flow occurring along $30^\circ$.
We plot
$\left\langle v_x\right\rangle$,
$\left\langle v_y\right\rangle$, and $\theta_\mathrm{sk}$ versus $j$
for the dimer-trimer
$N_\mathrm{sk}/N_m = 5/2$ filling in
Fig.~\ref{fig:7}(c, d).
The transport signatures have features very
similar to those that appear at the dimer
and trimmer fillings.
In Fig.~\ref{fig:8}(f) we show the trajectory of a
single skyrmion at $N_\mathrm{sk}/N_m = 5/2$
for $j = 5\times10^{10}$ A m $^{-2}$,
where the flow is disordered.

\section{Particle-Based Model}

\begin{figure}
  \centering
  \includegraphics[width=0.7\columnwidth]{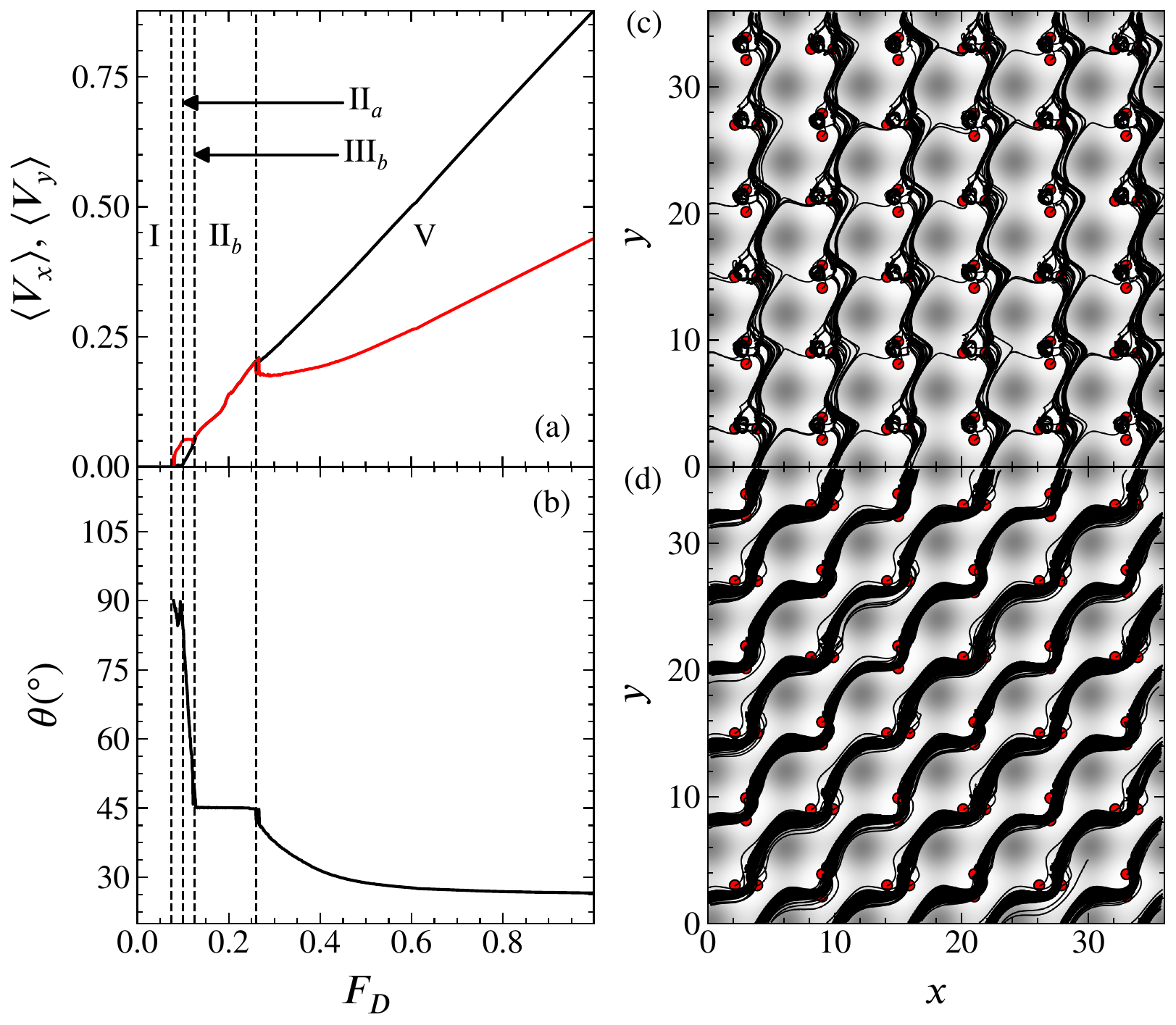}
  \caption{(a) Average skyrmion velocities
    $\langle V_x\rangle$ (black) and $\langle V_y\rangle$ (red) and
    (b) the skyrmion Hall angle $\theta$ vs $F_D$ from a particle-based
    model at a dimer filling of
    $N_\mathrm{sk}/N_m = 2.0$ on a square substrate.
    I: pinned phase.
    II$_a$: 90$^\circ$ transverse motion.
    II$_b$: 45$^\circ$ directionally locked motion.
    III$_b$: transitional regime between phases II$_a$ and II$_b$.
    V: disordered motion.
    (c) Skyrmion trajectories in phase II$_a$ at $F_D=0.085$.
    (d) Skyrmion trajectories in phase II$_b$ at $F_D=0.19$.
    Animations showing the phase II$_a$ and II$_b$ motion are available
    in the supplemental material \cite{suppl}.
  }
  \label{fig:9}
\end{figure}

We next turn to results from a particle-based model. As described in the Methods
section,
we match the parameters of the particle-based simulations
to those of the atomistic simulations.
We first focus on the $N_\mathrm{sk}/N_m = 2.0$ filling
for the square substrate, which forms the dimer state shown in
Fig.~\ref{fig:1}(a).
In Fig.~\ref{fig:9}(a,b) we plot
$\left\langle V_x\right\rangle$, $\left\langle V_y\right\rangle$,
and $\theta$ versus $F_D$.
A pinned phase I appears for $F_D < 0.075$,
while we observe the $90^\circ$ transverse
motion phase II$_a$ for $0.075 < F_D < 0.1$.
In Fig.~\ref{fig:9}(c)
we show the skyrmion trajectories at $F_D=0.085$ in phase II$_a$.
There is occasional hopping of individual skyrmions along the
$x$ direction, but the majority of the motion is locked
along the $y$ direction.
Over the range $0.125 < F_D < 0.26$,
the 45$^\circ$ directional locking phase II$_b$ appears.
The motion is solitonic,
as shown Fig.~\ref{fig:9}(d) at $F_D=0.19$.
In between phases II$_a$ and II$_b$, over the range
$0.1 < F_D < 0.125$, there is a transitional state
labeled III$_b$
in which there is a combination of transverse
motion with an increased amount of hopping along the $x$ direction.
At higher drives of $F_D>0.26$, the system
enters the more disordered phase V, there
is no directional locking,
and the skyrmion Hall angle
approaches the intrinsic value.
A drop in $\left\langle V_y\right\rangle$ corresponding to
negative differential conductivity
occurs at the transition between phase II$_b$ and phase V.

\begin{figure}
  \centering
  \includegraphics[width=0.7\columnwidth]{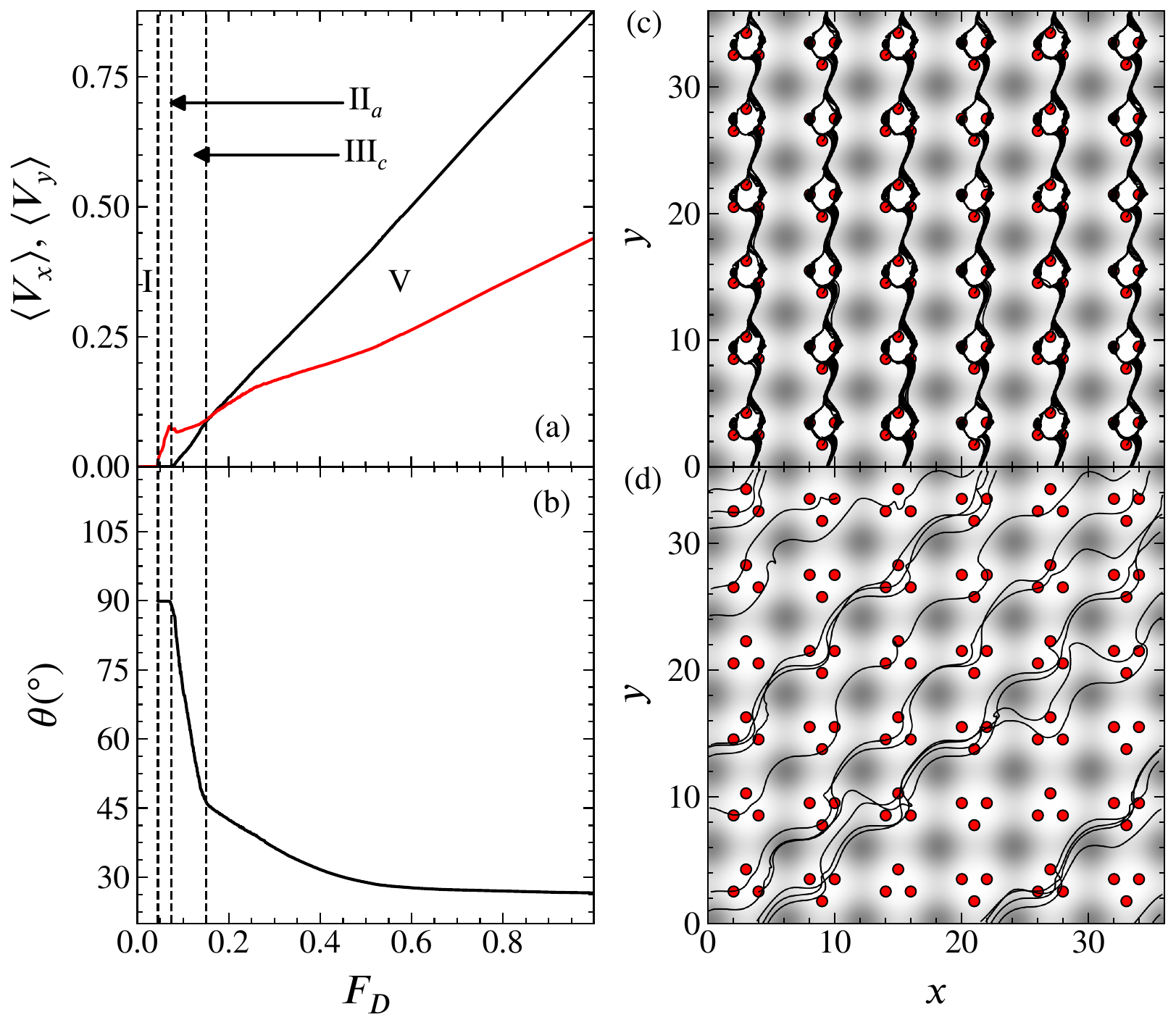}
  \caption{Average skyrmion velocities
    $\langle V_x\rangle$ (black) and $\langle V_y\rangle$ (red)
    and (b) the skyrmion Hall angle $\theta$ vs $F_D$ from a
    particle-based model at a trimer filling
    of $N_\mathrm{sk}/N_m = 3.0$ on a square substrate.
    I: pinned phase.
    II$_a$: 90$^\circ$ transverse motion.
    III$_c$: transitional regime between phases II$_a$ and V.
    V: disordered motion.
    (c) Skyrmion trajectories in phase II$_a$ at $F_D=0.05$.
    II$_a$.
    (d) Trajectory of a single skyrmion in phase V at $F_D=0.2$.
    Animations showing the phase II$_a$ and V motion are available
    in the supplemental material \cite{suppl}.
  }
  \label{fig:10}
\end{figure}

\begin{figure}
  \centering
  \includegraphics[width=0.7\columnwidth]{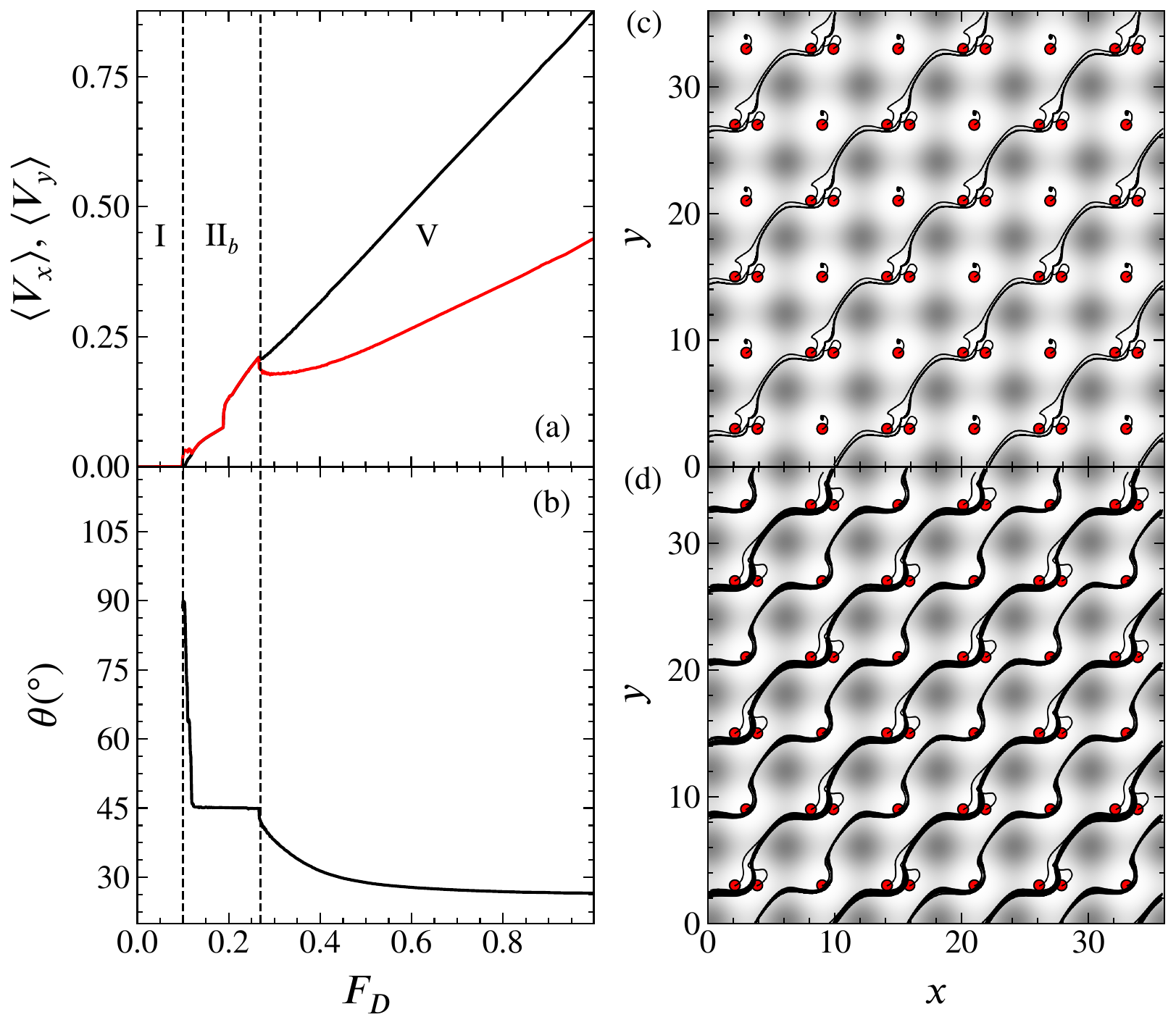}
  \caption{(a) Average skyrmion velocities
    $\langle V_x\rangle$ (black) and $\langle V_y\rangle$ (red) and
    (b) the skyrmion Hall angle $\theta$ vs $F_D$ from a particle-based
    model at a monomer-dimer filling of
    $N_\mathrm{sk}/N_m = 3/2$ on a square substrate.
    I: pinned phase.
    II$_b$: 45$^\circ$ directionally locked motion.
    V: disordered motion.
    (c) Skyrmion trajectories in phase II$_b$ at $F_D=0.15$.
    (d) Trajectory of a single skyrmion in phase II$_b$ at $F_D=0.2$
    where all of the skyrmions are moving.
    Animations showing the phase II$_b$ motion at different
    $F_D$ values are available
    in the supplemental material \cite{suppl}.
  }
  \label{fig:11}
\end{figure}

In Fig.~\ref{fig:10}(a) we plot
$\left\langle V_x\right\rangle$, $\left\langle V_y\right\rangle$,
and $\theta$ versus $F_{D}$ for the trimer state
at $N_\mathrm{sk}/N_m = 3.0$ on a square substrate.
The pinned phase I depins into the
transverse motion phase II$_a$.
We illustrate
the skyrmion trajectories in phase II$_a$ at $F_D=0.05$ in
Fig.~\ref{fig:10}(c), which shows that
the transverse flow occurs via the motion of a
soliton.
As the drive increases we observe a transitional state,
labeled III$_c$, between phase II$_a$ and the disordered
phase V flow that appears at higher drives.
Unlike the dimer lattice state, we observe
no phase II$_b$ or $45^\circ$ directionally locked flow.
This is due to the creation of disordered flow structures that make
the motion more thermal in character and prevent ordered
channels from forming along the $45^\circ$ direction.
For increasing drive in phase V,
the motion gradually rotates into the direction of
the intrinsic Hall angle.
In the particle-based model,
all of the skyrmions within each trimer participate in the
phase II$_a$ motion. This
is in contrast to the atomistic model results from Fig.~\ref{fig:3}(c),
where one skyrmion in each trimer remained stationary throughout
the phase II$_a$ regime. The difference arises because the individual
skyrmions can distort in the atomistic model but have fixed effective
sizes in the particle-based model.

\begin{figure}
  \centering
  \includegraphics[width=0.7\columnwidth]{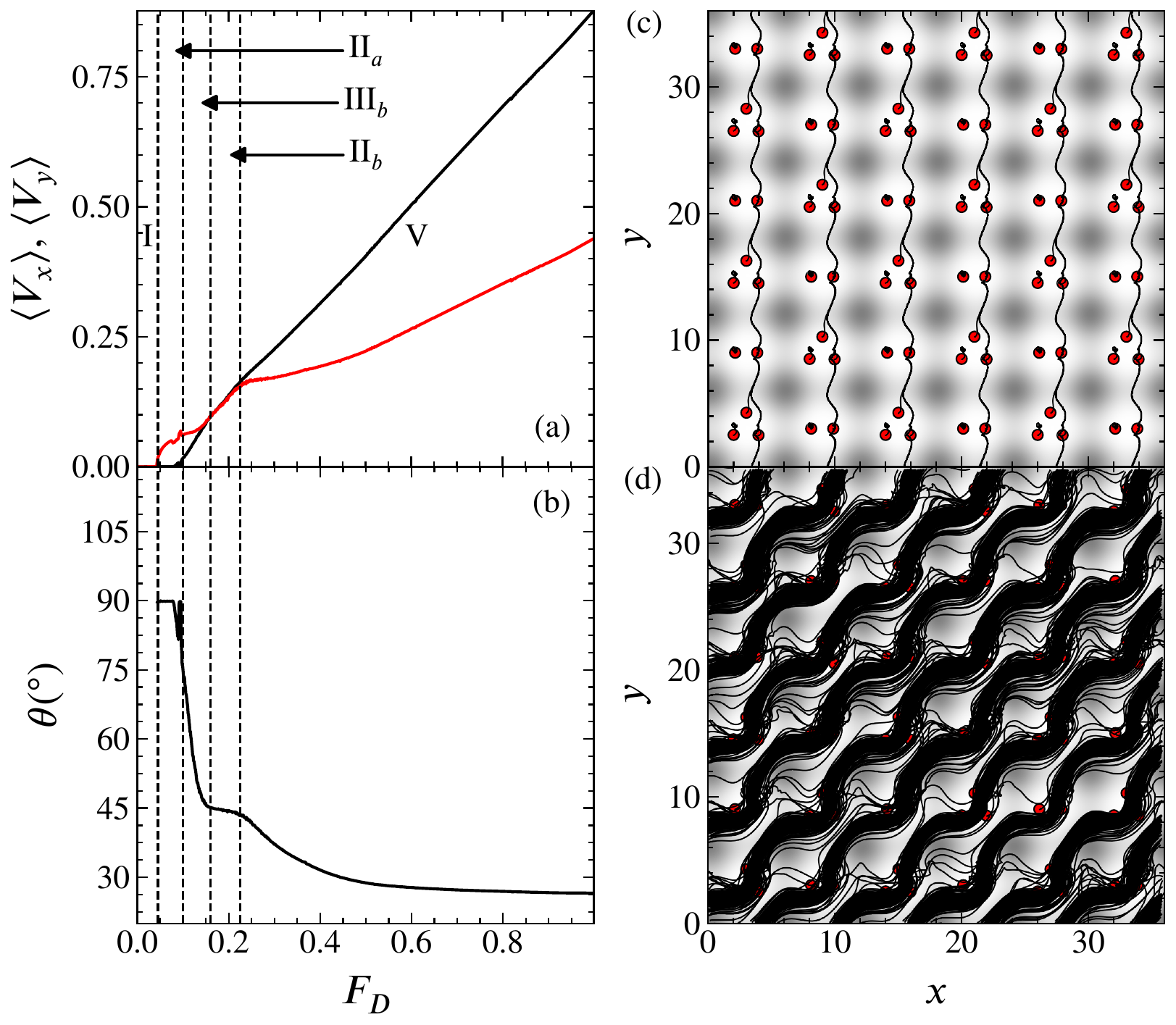}
  \caption{(a) Average skyrmion velocities $\langle V_x\rangle$
    (black) and $\langle V_y\rangle$ (red) and
    (b) the skyrmion Hall angle $\theta$ vs $F_D$ from a particle-based
    model at a dimer-trimer filling of
    $N_\mathrm{sk}/N_m = 5/2$ on a square substrate.
    I: pinned phase.
    II$_a$: 90$^\circ$ transverse motion.
    II$_b$: 45$^\circ$ directionally locked motion.
    III$_b$: transitional regime between phases II$_a$ and II$_b$.
    V: disordered motion.
    (c) Skyrmion trajectories in phase II$_a$ at $F_D=0.05$.
    (d) Trajectory of a single skyrmion in phase II$_b$ at $F_D=0.2$.
    Animations showing the phase II$_a$ and II$_b$ motion are available
    in the supplemental material \cite{suppl}.
  }
  \label{fig:12}
\end{figure}

We plot
$\left\langle V_x\right\rangle$, $\left\langle V_y\right\rangle$, and
$\theta$ versus $F_D$
in Fig.~\ref{fig:11}(a, b)
for a monomer-dimer filling of $N_\mathrm{sk}/N_m = 3/2$ on a square substrate.
A pinned phase I appears for $F_D < 0.1$,
and there is a 45$^\circ$ directional locking phase II$_b$
over the region $0.1 < F_D < 0.27$.
The directionally locked phase II$_b$ contains an internal
depinning transition. For lower drives,
soliton motion occurs only along diagonal channels that contain dimers,
while channels containing monomers remain pinned, as illustrated
in
Fig.~\ref{fig:11}(c) at $F_D=0.15$.
For higher drives, the monomers can also participate in the soliton
motion, as shown
in Fig.~\ref{fig:11}(d) at $F_D=0.2$.
At the transition from phase II$_b$ to the high drive
disordered phase V flow,
there is a drop in the skyrmion velocity.

\begin{figure}
  \centering
  \includegraphics[width=0.7\columnwidth]{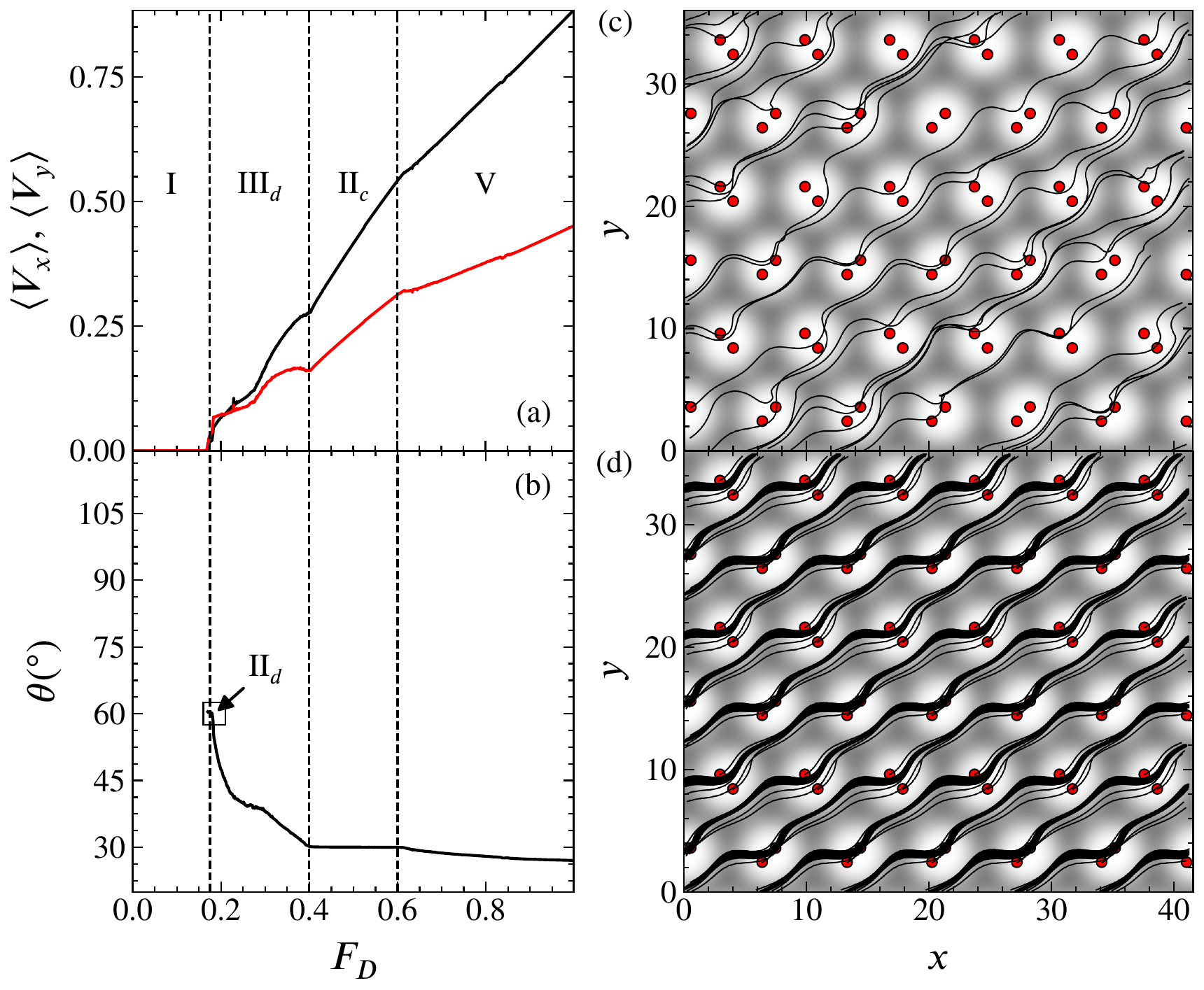}
  \caption{(a) Average skyrmion velocities $\langle V_x\rangle$
    (black) and $\langle V_y\rangle$ (red) and
    (b) the skyrmion Hall angle $\theta$ vs $F_D$
    from a particle-based model at a dimer filling
    of $N_\mathrm{sk}/N_m = 2.0$
    on a triangular substrate.
    I: pinned phase.
    II$_c$: 30$^\circ$ directionally locked motion.
    II$_d$: 60$^\circ$ directionally locked motion.
    III$_d$: transitional regime between phases II$_d$ and II$_c$.
    V: disordered motion.
    (c) Trajectory of a single skyrmion
    in phase III$_d$ at $F_D=0.35$.
    (d) Skyrmion trajectories
    in phase II$_c$ at $F_D=0.5$.
    Animations showing the phase III$_d$ and II$_c$ motion are available
    in the supplemental material \cite{suppl}.
  }
  \label{fig:13}
\end{figure}

For the dimer-trimer filling
of the square substrate at $N_\mathrm{sk}/N_m = 5/2$,
Fig.~\ref{fig:12}(a, b) shows
$\left\langle V_x\right\rangle$, $\left\langle V_y\right\rangle$,
and $\theta_\mathrm{sk}$ versus $F_D$, where
phases I, II$_a$, III$_b$, II$_b$, and V appear.
The transverse motion
in phase II$_a$ occurs through soliton flow,
as shown in Fig.~\ref{fig:12}(c) at
$F_D=0.05$.
In the directionally locked
phase II$_b$,
the skyrmions move preferentially along a 45$^\circ$ path with
some occasional hopping along the $x$ direction,
as illustrated in Fig.~\ref{fig:12}(d) at $F_D=0.2$.

Figure~\ref{fig:13}(a, b) shows
$\left\langle V_x\right\rangle$, $\left\langle V_y\right\rangle$,
and $\theta$ versus $F_D$ for the dimer filling of
$N_\mathrm{sk}/N_m = 2.0$ on a triangular substrate.
From the pinned phase I, no transverse motion occurs at depinning,
and the initial flow is of solitons moving along
$60^\circ$, labeled phase II$_d$. This phase II$_d$ is very narrow and
although it also exists in Fig.~\ref{fig:13}(a), it is only
highlighted in Fig.~\ref{fig:13}(b).
As $F_D$ increases, we find
region III$_d$ flow, which is transitional between regions II$_d$ and
II$_c$.
In region III$_d$, illustrated in Fig.~\ref{fig:13}(c) at $F_D=0.35$, $\theta$
decreases with increasing $F_D$
until there is a directional locking along $30^\circ$
and the system enters phase
II$_c$.
A small dip in the velocities appears at the transition between
phases
III$_d$ and II$_c$.
The velocities increase linearly with increasing drive
in phase II$_c$ and the flow is ordered,
as shown in Fig.~\ref{fig:13}(d) at $F_D=0.5$.
At
higher drives, the system transitions to phase V, where there is no
directional locking and the flow is disordered.

\begin{figure}
  \centering
  \includegraphics[width=0.7\columnwidth]{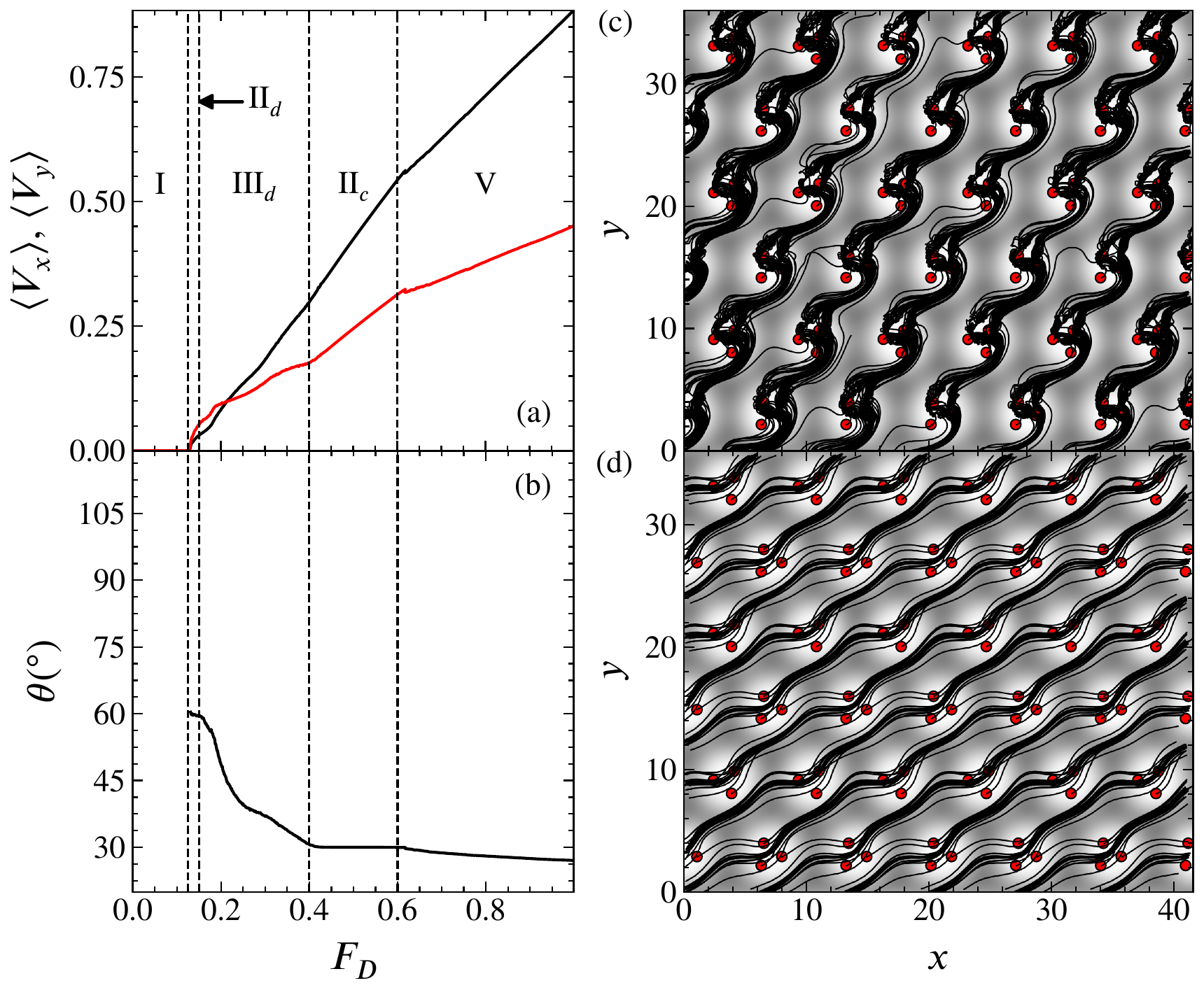}
  \caption{(a) Average skyrmion velocities
    $\langle V_x\rangle$ (black) and $\langle V_y\rangle$ (red) and
    (b) the skyrmion Hall angle $\theta$ vs $F_D$ from a particle-based
    model at a trimer filling of
    $N_\mathrm{sk}/N_m = 3.0$
    on a triangular substrate.
    I: pinned phase.
    II$_d$: 60$^\circ$ directionally locked motion.
    II$_c$: 30$^\circ$ directionally locked motion.
    III$_d$: transitional regime between phases II$_d$ and II$_c$.
    V: disordered motion.
    (c) Skyrmion trajectories in phase II$_d$ at $F_D=0.14$.
    (d) Trajectory of a single skyrmion in phase II$_c$ at $F_D=0.5$.
    Animations showing the phase II$_d$ and II$_c$ motion are available
    in the supplemental material \cite{suppl}.
  }
  \label{fig:14}
\end{figure}

In Fig.~\ref{fig:14}(a, b) we plot
$\left\langle V_x\right\rangle$, $\left\langle V_y\right\rangle$,
and $\theta$ versus $F_D$ for a trimer filling of
$N_\mathrm{sk}/N_m = 3.0$.
on a triangular substrate.
We again observe phases I, II$_d$, III$_d$, II$_c$, and V, and
find no transverse $90^\circ$
motion.
When the system depins into the locked phase II$_d$ flow,
highlighted in Fig.~\ref{fig:14}(b) at $F_D=0.14$,
$\theta=60^\circ$ and the motion is almost entirely along the
major symmetry direction of the substrate, with occasional hopping
in the $x$ direction.
As $F_D$ increases, the system enters the transitional region III$_d$ and
the skyrmion Hall angle $\theta$ decreases until it
becomes locked to $30^\circ$ at the transition into
phase II$_c$.
The trajectories at $F_D = 0.5$, illustrated in Fig.~\ref{fig:14}(d),
fall in the center of the phase II$_c$ range which extends from
$0.4 < F_D < 0.6$, and show
channels of motion along 30$^\circ$ with
occasional hopping between adjacent wells.
At high drives in phase V, the flow is disordered.

\begin{figure}
  \centering
  \includegraphics[width=0.7\columnwidth]{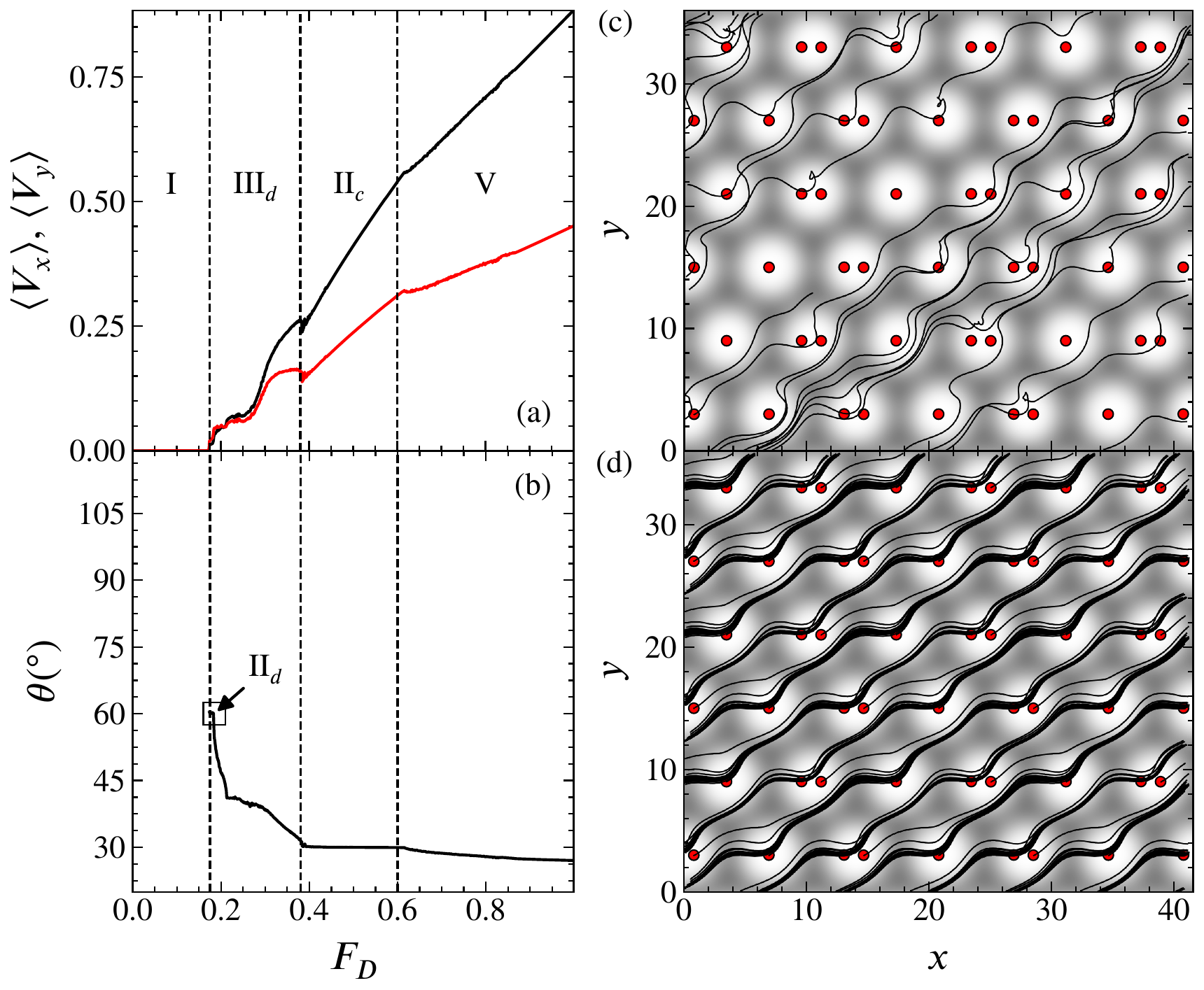}
  \caption{(a) Average skyrmion velocities $\langle V_x\rangle$
    (black) and $\langle V_y\rangle$ (red) and
    (b) the skyrmion Hall angle $\theta$ vs $F_D$ from a
    particle-based model at a monomer-dimer filling of
    $N_\mathrm{sk}/N_m = 3/2$
    on a triangular substrate.
    I: pinned phase.
    II$_d$: 60$^\circ$ directionally locked motion.
    II$_c$: 30$^\circ$ directionally locked motion.
    III$_d$: transitional regime between phases II$_d$ and II$_c$.
    V: disordered motion.
    (c) Trajectory of a single skyrmion
    in phase III$_d$ at $F_D=0.3$.
    (d) Skyrmion trajectories in phase II$_c$ at $F_D=0.5$.
    Animations showing the phase III$_d$ and II$_c$ motion are available
    in the supplemental material \cite{suppl}.
  }
  \label{fig:15}
\end{figure}

The monomer-dimer filling of
$N_\mathrm{sk}/N_m = 3/2$ for a triangular substrate is shown
in Fig.~\ref{fig:15}(a, b), where we plot
$\left\langle V_x\right\rangle$, $\left\langle V_y\right\rangle$,
and $\theta_\mathrm{sk}$ versus $F_D$.
The phases that appear are similar to those found for the
$N_\mathrm{sk}/N_m=2.0$ dimer filling, and there is
a strong drop
in the velocity at the III$_d$ to II$_c$ transition.
Fig.~\ref{fig:15}(c) shows
the trajectory of a single skyrmion in phase III$_d$ at $F_{D}= 0.3$,
where the motion is not locked to the substrate but smoothly
changes in angle as the drive increases.
In Fig.~\ref{fig:15}(d) we show trajectories for all
of the skyrmions in phase II$_c$ at $F_D=0.5$,
where the motion is almost entirely locked along $30^\circ$ with
occasional hops in the $x$ direction. The directional locking is not
quite as strong as what we observe for the atomistic model in
Fig.~\ref{fig:7}(e), where no $x$ direction hopping occurs, and this
is due to the inability of the particle-based skyrmions to deform.

\begin{figure}
  \centering
  \includegraphics[width=0.7\columnwidth]{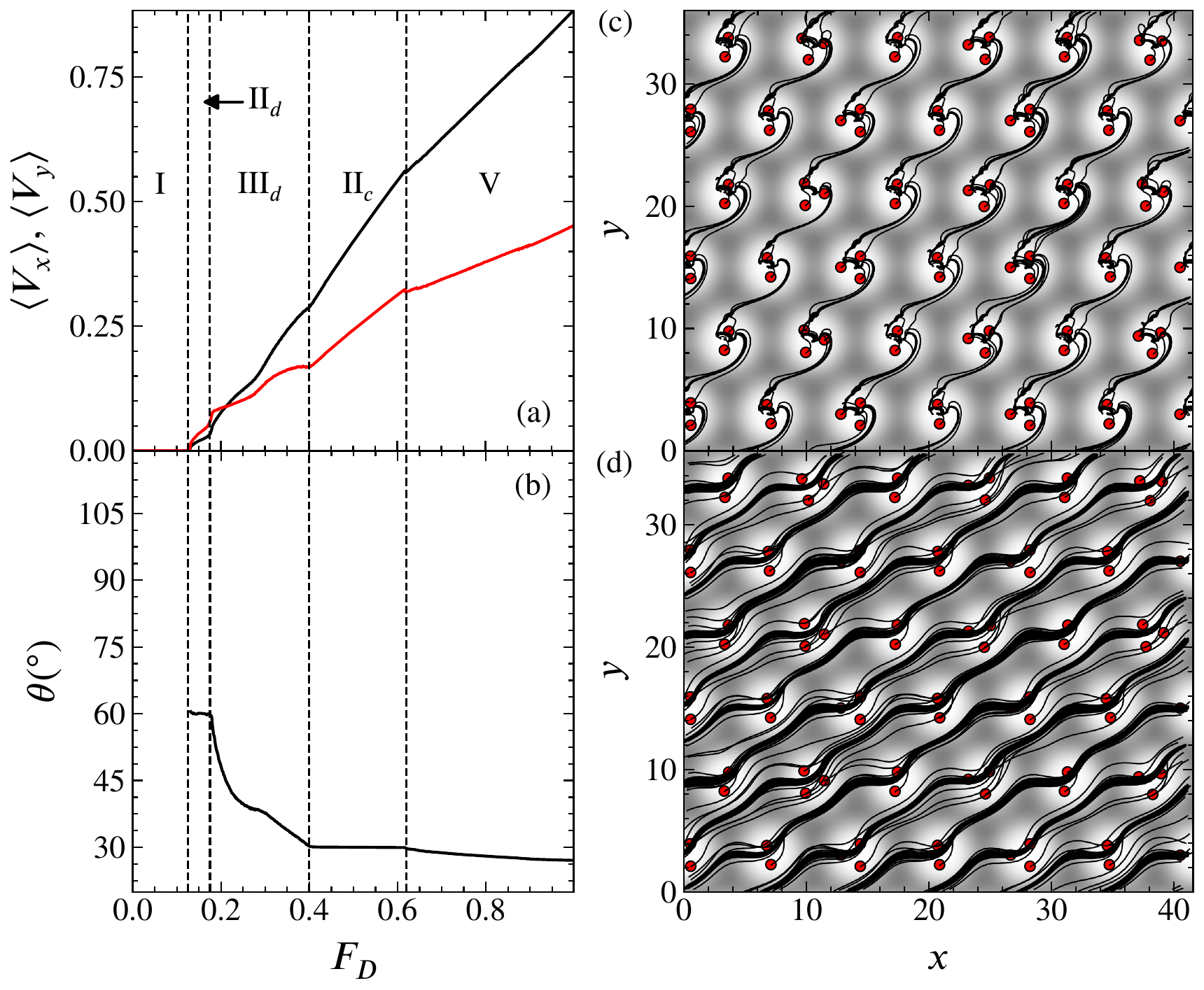}
  \caption{(a) Average skyrmion velocities $\langle V_x\rangle$
    (black) and $\langle V_y\rangle$ (red) and
    (b) the skyrmion Hall angle $\theta$ vs $F_D$ from a
    particle-based model at a dimer-trimer filling of
    $N_\mathrm{sk}/N_m = 5/2$
    on a triangular substrate.
    I: pinned phase.
    II$_d$: 60$^\circ$ directionally locked motion.
    II$_c$: 30$^\circ$ directionally locked motion.
    III$_d$: transitional regime between phases II$_d$ and II$_c$.
    V: disordered motion.
    (c) Skyrmion trajectories in phase II$_d$ at $F_D=0.15$.
    (d) Skyrmion trajectories in phase II$_c$ at $F_D=0.5$.
    Animations showing the phase II$_d$ and II$_c$ motion are available
    in the supplemental material \cite{suppl}.
  }
  \label{fig:16}
\end{figure}

In Fig.~\ref{fig:16}(a, b) we plot
$\left\langle V_x\right\rangle$, $\left\langle V_y\right\rangle$,
and $\theta_\mathrm{sk}$ versus $F_D$
for a triangular substrate at the dimer-trimer
filling of $N_\mathrm{sk}/N_m = 5/2$.
The directionally locked phase II$_d$, in which the skyrmions move along
$60^\circ$, that appears just above depinning is much more extended
compared to the fillings that have been described above, and is
illustrated at $F_D=0.15$ in Fig.~\ref{fig:16}(c).
After the system passes through a transitional III$_d$ state,
a $30^\circ$ directionally locked phase II$_c$ appears which has the
motion shown for a drive of $F_D=0.5$ in Fig.~\ref{fig:16}(d).
At higher drives, the system enters the disordered flow region V state.

In general, the particle model qualitatively shows
the same
phases as the atomistic models,
including transverse motion, soliton flows, directional locking,
and the disordered phases; however, there are several
quantitative differences between the two models.
This could be due to the ability of the
skyrmions to distort or change size as they move,
which is present in the atomistic model
but not in the particle-based model. These distortions and size
changes introduce
additional fluctuations so that
the robustness of directional locking to
the higher order symmetry locking directions is reduced.
For both models, we find strong solitonic motion in the
mixed molecular crystal states,
when a portion of the skyrmions are pinned and kink
motion can occur.
This is due to the strong difference in the effectiveness of
the pinning for monomers compared to dimers. Since the monomers can sit
at the bottom of the pinning well, they are much more difficult to depin
than individual skyrmions of a dimer.

\section{Varied Anisotropy}

\begin{figure}
  \centering
  \includegraphics[width=0.7\columnwidth]{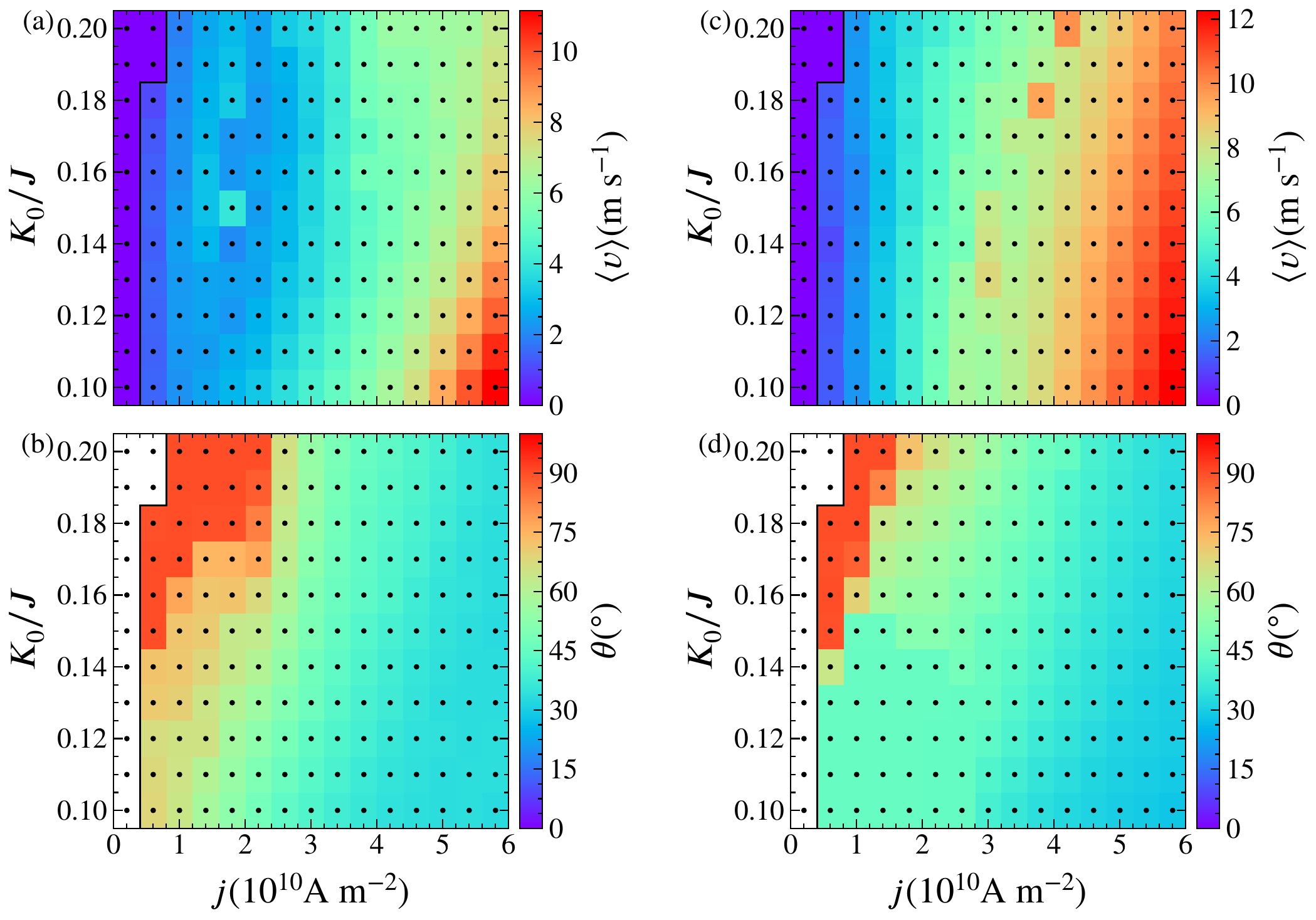}
  \caption{Heat maps
    as a function of $K_0/J$ vs $j$ for atomistic simulations on
    a square substrate.
    (a, c) The absolute skyrmion velocity
    $\langle v\rangle=\sqrt{\langle v_x\rangle^2+\langle v_y^2\rangle}$.
    (b, d) The Hall angle $\theta$.
    (a, b) The dimer filling at
    $N_\mathrm{sk}/N_m = 2.0$.
    (c, d) The trimer filling at $N_\mathrm{sk}/N_m = 3.0$.
    For each filling,
    there is a critical
    anisotropy level $K_0/J$ that must be exceeded
    for transverse motion to occur.
  }
  \label{fig:17}
\end{figure}

We next consider the evolution of the phases over a range of
anisotropies to show that the results are robust for the atomistic
model. In Fig.~\ref{fig:17}(a, b), we plot heat maps of the net
skyrmion velocity $\langle v\rangle$ and the Hall angle
$\theta$ as a function of $K_0/J$ versus $j$ for a square
substrate at the dimer filling of $N_\mathrm{sk}/N_m = 2.0$.
For $K_0/J > 0.15$, there is
a region of transverse motion where the Hall angle is nearly
$90^\circ$. The pinned region with $\langle v\rangle=0$
and the transverse flow region with $\theta=90^\circ$ both grow in
extent as $K_0/J$ increases.
This indicates that a critical level of
anisotropy is required to permit transverse motion to
occur. Figure~\ref{fig:17}(c, d) shows $\langle v\rangle$ and $\theta$
heat maps for the same system at a trimer filling of
$N_\mathrm{sk}/N_m = 3.0$.
The overall trend of the phases is similar to the dimer filling,
and transverse motion only occurs
when $K_0/J > 0.15$.
The trimer filling in
Fig.~\ref{fig:17}(c) exhibits
some instances of
negative differential mobility, where the velocity
is greatest at intermediate drives.

\begin{figure}
  \centering
  \includegraphics[width=0.7\columnwidth]{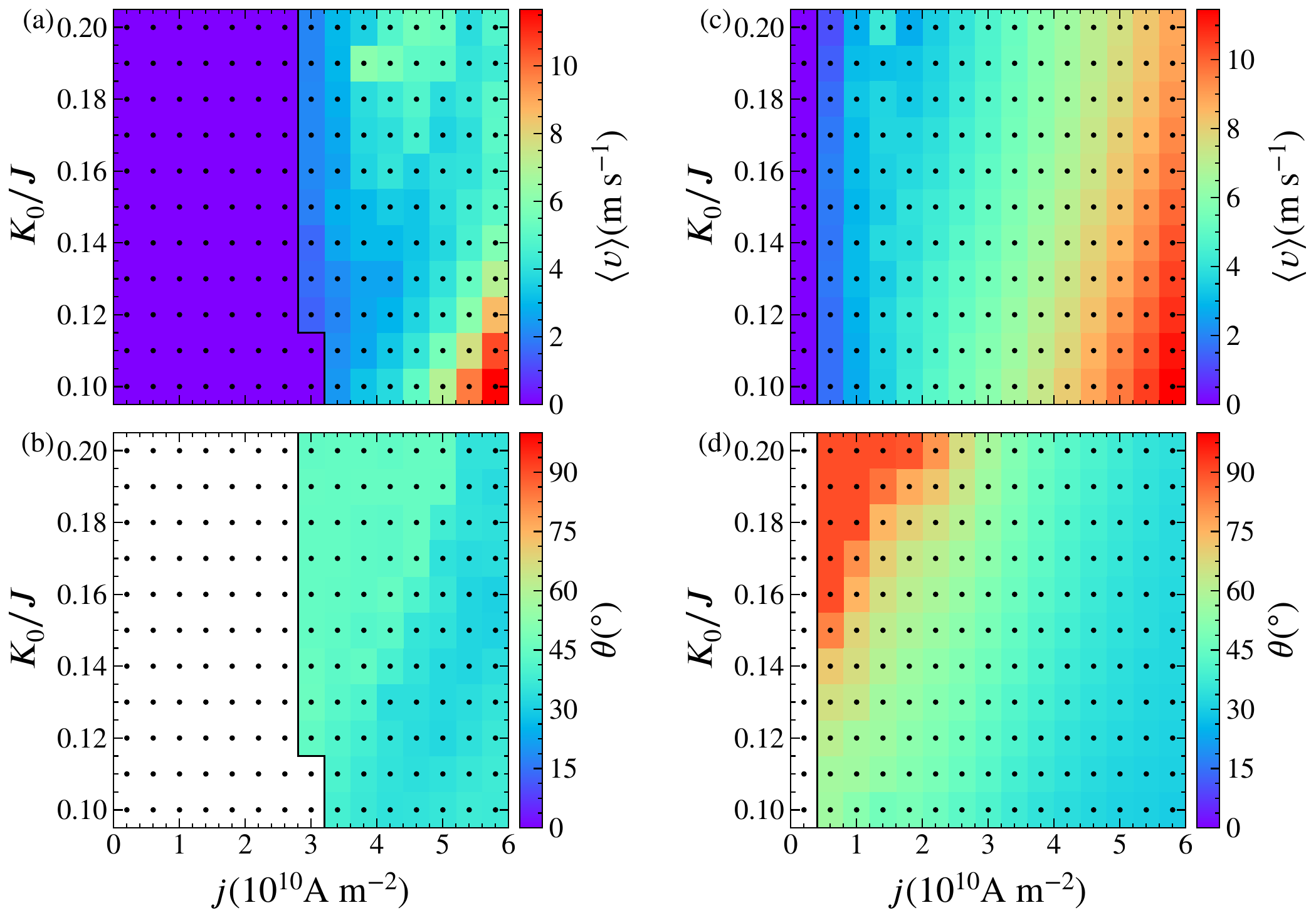}
  \caption{Heat maps
    as a function of $K_0/J$ vs $j$ for atomistic simulations on
    a square substrate.
    (a, c) The absolute skyrmion velocity
    $\langle v\rangle=\sqrt{\langle v_x\rangle^2+\langle v_y^2\rangle}$.
    (b, d) The Hall angle $\theta$.
    (a, b) The monomer-dimer filling at
    $N_\mathrm{sk}/N_m = 3/2$, where there is an extensive pinned region
    and no transverse motion regime.
    (c, d) The dimer-trimer filling at $N_\mathrm{sk}/N_m = 5/2$.
    For each filling,
    there is a critical
    anisotropy level $K_0/J$ that must be exceeded
    for transverse motion to occur.
  }
  \label{fig:18}
\end{figure}

In Fig.~\ref{fig:18}(a, b), we show heat maps of $\langle v\rangle$
and $\theta$ as a function of
$K_0/J$ versus $j$ for a square
substrate at
the monomer-dimer filling of $N_\mathrm{sk}/N_m = 3/2$.
Here, the pinned region is very large and extends up to
$j = 3.2\times10^{10}$ A m$^{-2}$.
Additionally, there is no transverse motion
regime, and the Hall angle remains close to
$\theta=30^\circ$ for all moving states.
The extended pinned regime arises because the
monomers are much more strongly pinned than the dimers.
Since the
skyrmions have a checkerboard ordering at this filling, when dimers
attempt to depin by moving
in the transverse direction, they are blocked by the more
strongly pinned monomers.
The $\langle v\rangle$ and $\theta$ heat maps
as a function of $K_0/J$ versus $j$ shown
in Fig.~\ref{fig:18}(c, d) for the dimer-trimer filling of
$N_\mathrm{sk}/N_m = 5/2$ indicate that
the pinned regime has become much smaller,
and a regime of transverse motion
emerges when $K_0/J > 0.16$.

\begin{figure}
  \centering
  \includegraphics[width=0.7\columnwidth]{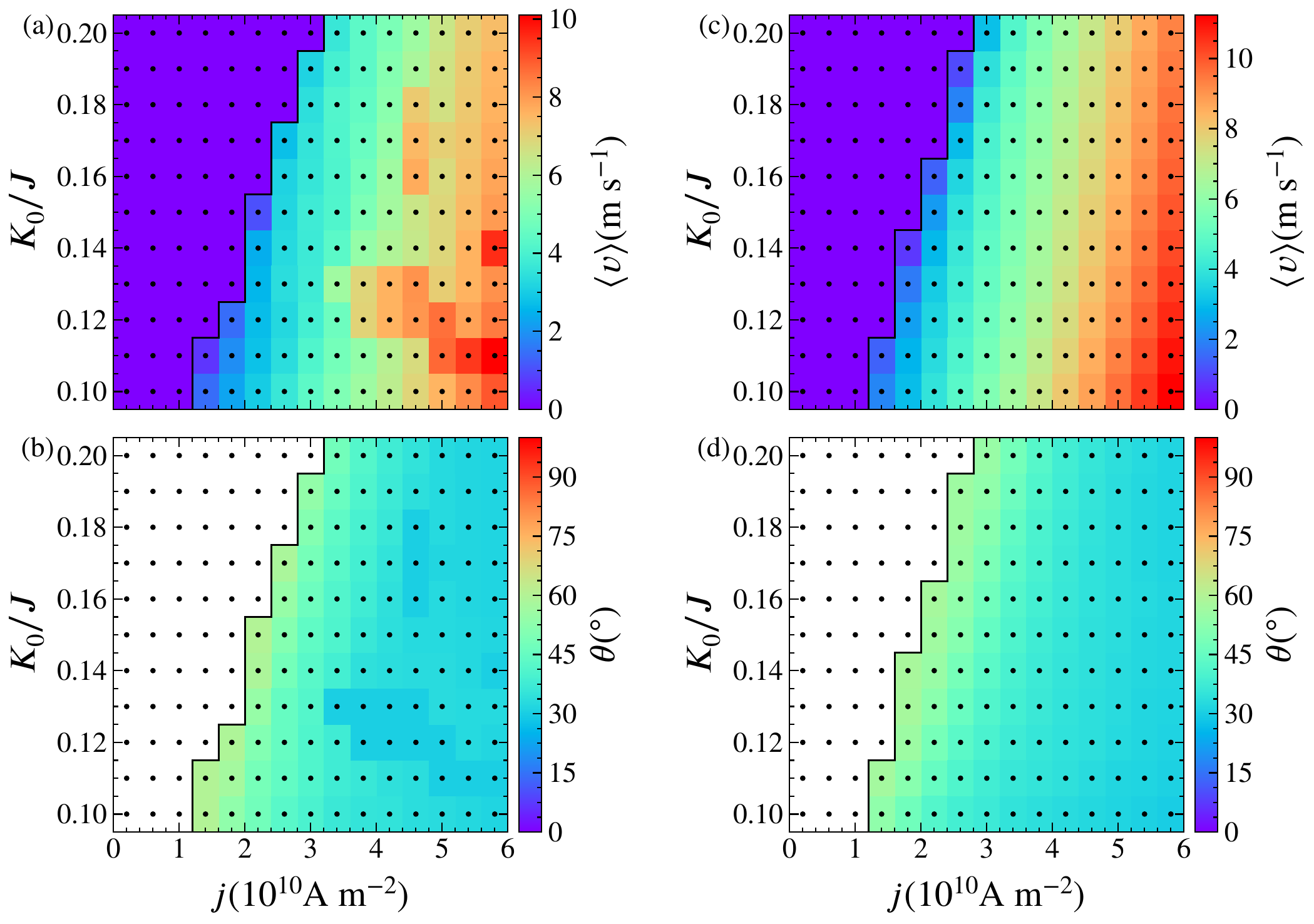}
  \caption{Heat maps
    as a function of $K_0/J$ vs $j$ for atomistic simulations on
    a triangular substrate.
    (a, c) The absolute skyrmion velocity
    $\langle v\rangle=\sqrt{\langle v_x\rangle^2+\langle v_y^2\rangle}$.
    (b, d) The Hall angle $\theta$.
    (a, b) The dimer filling at
    $N_\mathrm{sk}/N_m = 2.0$.
    (c, d) The trimer filling at $N_\mathrm{sk}/N_m = 3.0$.
    Unlike the case of dimers and trimers on a square substrate,
    for the triangular substrate no transverse motion regime appears.
  }
  \label{fig:19}
\end{figure}

For atomistic simulations of the triangular substrate,
in Fig.~\ref{fig:19}(a, b) we plot heat maps of $\langle v\rangle$
and $\theta$ as a function of $K_0/J$ versus $j$ at the
dimer filling of
$N_\mathrm{sk}/N_m = 2.0$.
The pinned region is much
larger than for the dimer state on a square substrate, and
increases linearly in width
with increasing $K_0/J$.
The enhanced pinning arises because
the drive is not aligned with one of the
major symmetry directions of the triangular substrate,
such as $60^\circ$, whereas for the
square substrate, both the current and the drive are aligned with
major symmetry directions of $90^\circ$ and 0$^\circ$.
When depinning occurs on the triangular substrate, the Hall
angle is close to $60^\circ$,
and $\theta$ decreases toward the intrinsic
Hall angle value as $j$
increases.
There are several windows of directional locking to
$30^\circ$, which appear as non-monotonic changes in velocity and
a plateau of the Hall angle at
$30^\circ$.
For the trimer
filling of
$N_\mathrm{sk}/N_m = 3.0$ on the triangular lattice,
the heat maps of $\langle v\rangle$ and $\theta$ as a function of
$K_0/J$ versus $j$ in
Fig.~\ref{fig:19}(c, d) indicate that
there is again a strong pinning regime
that increases in width with increasing $K_0/J$,
and the Hall angle near depinning is close to $60^\circ$.
The trimer filling shows a
smoother decrease in the Hall angle with increasing drive
compared to the dimer filling.

\begin{figure}
  \centering
  \includegraphics[width=0.7\columnwidth]{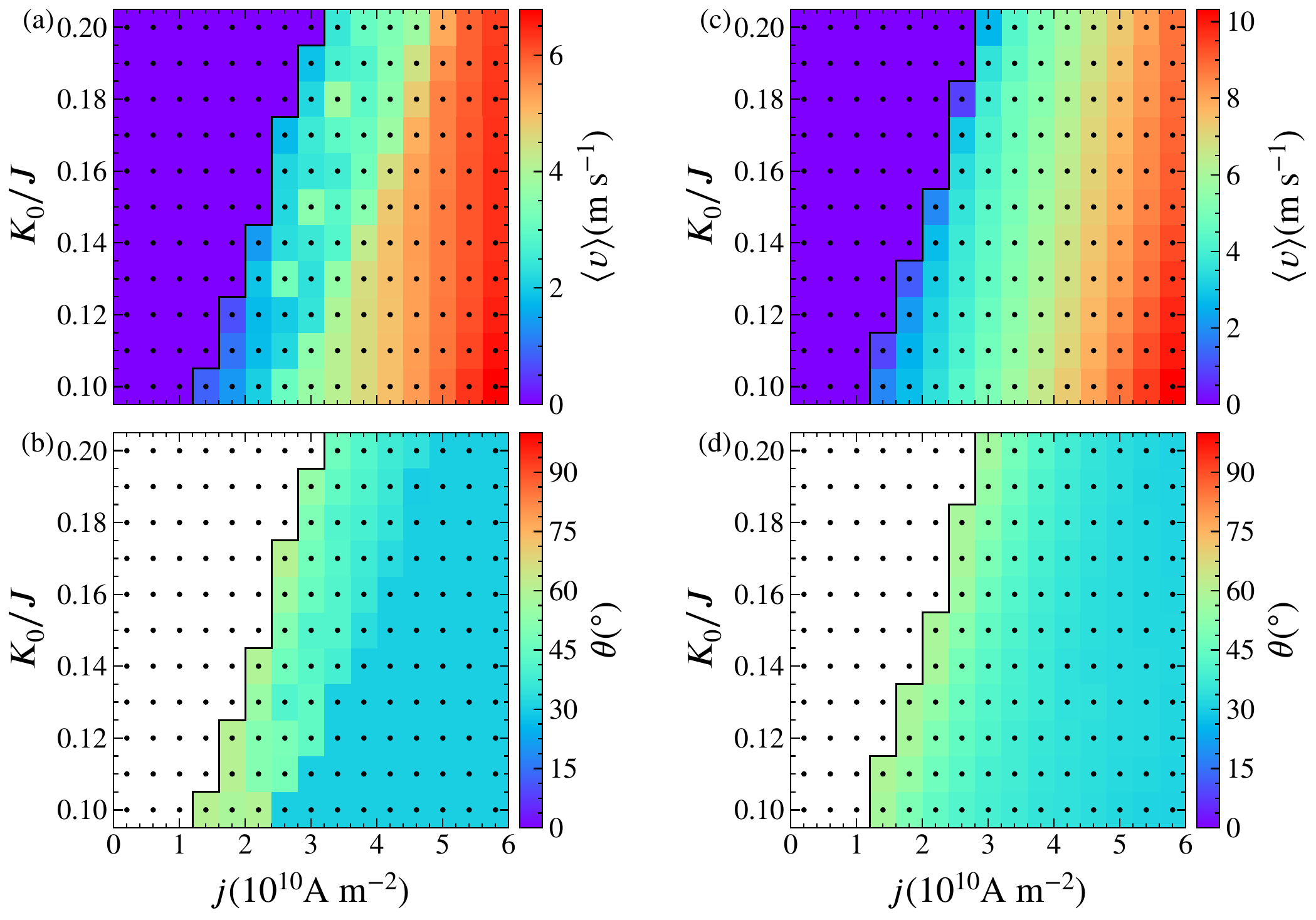}
  \caption{Heat maps
    as a function of $K_0/J$ vs $j$ for atomistic simulations on
    a triangular substrate.
    (a, c) The absolute skyrmion velocity
    $\langle v\rangle=\sqrt{\langle v_x\rangle^2+\langle v_y^2\rangle}$.
    (b, d) The Hall angle $\theta$.
    (a, b) The monomer-dimer filling at
    $N_\mathrm{sk}/N_m = 3/2$.
    (c, d) The dimer-trimer filling at $N_\mathrm{sk}/N_m = 5/2$.
  }
  \label{fig:20}
\end{figure}

In Fig.~\ref{fig:20}(a, b), we plot heat maps
of $\langle v\rangle$ and $\theta$ as a function of
$K_0/J$ versus $j$
for atomistic simulations of a triangular substrate
at the monomer-dimer filling of $N_\mathrm{sk}/N_m = 3/2$.
Here an extended pinning
phase is present due to the strongly pinned monomers.
The heat maps
of $\langle v\rangle$ and $\theta$ as a function of
$K_0/J$ versus $j$
for the dimer-trimer filling of
$N_\mathrm{sk}/N_m=5/2$ in the same
sample shown in Fig.~\ref{fig:20}(c, d) indicate that
the Hall angle is close to $60^\circ$
above depinning and decreases with increasing drive to the intrinsic
value.
The heat maps for both the square and triangular substrates for
varied anisotropy $K_0/J$ indicate that the results described in this
work should remain robust over a wide range of
parameters.

\section{Summary}

We have investigated the driven dynamics of skyrmion molecular
crystals using atomistic and particle-based simulations
of square and triangular substrates at
fillings of $N_\mathrm{sk}/N_m = 3/2$, 2.0, 5/2, and
$3.0$, where the skyrmions form dimers,
trimers, or mixed lattices that have both positional and orientational
order. For a square substrate, for several of the fillings the skyrmion
depinning occurs via the formation of running solitons
that travel transverse to the applied current.
At higher drives, the skyrmion Hall angle gradually
approaches its intrinsic value and can pass through multiple steps
and phases,
with negative differential conductivity appearing at some of the transitions
between flow states.
For the square substrate, directional locking in which the motion of the
skyrmions remains fixed to a constant direction over a window of applied
drives can occur along
$45^\circ$.
At the monomer-dimer filling of $N_\mathrm{sk}/N_m = 3/2$,
we observe a two-step depinning process in which
the dimers depin first and the monomers depin only at higher drives.
Due to the blocking of the flow by the monomers, there is no region of
transverse flow above the first depinning transition.
Triangular substrates do not support transverse flow due to their
symmetry, but can show
directional locking of the motion along $60^\circ$ and $30^\circ$.
Within a directionally locked state,
the flow is generally well ordered.
The pinned regime for the triangular substrate tends to be considerably
larger than that of the square substrate because
the drive is only aligned with a square substrate symmetry direction and is
not aligned with a triangular substrate symmetry direction.
The particle-based model produces
the same qualitative phases as the atomistic
model
but shows several quantitative differences.
We demonstrate that the phases we observe remain
robust over a range of anisotropy values, and that a
critical minimum anisotropy
level is required in order to permit
directional locking and transverse
flow phases to occur.

\section*{Acknowledgments}
These resources were funded by the Fundação de Amparo à Pesquisa do
Estado de São Paulo - FAPESP (Grant: 2021/04655-8). This work was
supported by the US Department of Energy through the Los Alamos
National Laboratory. Los Alamos National Laboratory is operated by
Triad National Security, LLC, for the National Nuclear Security
Administration of the U. S. Department of Energy (Contract
No. 892333218NCA000001).
J.C.B.S and N.P.V. acknowledge funding from Fundação de Amparo à
Pesquisa do Estado de São Paulo - FAPESP (Grants J.C.B.S 2023/17545-1
and 2022/14053-8, N.P.V 2024/13248-5).
We would like to thank FAPESP for providing the computational
resources used in this work (Grant: 2024/02941-1).

\bibliographystyle{unsrt}
\bibliography{mybib}

\end{document}